\documentclass[useAMS,usenatbib]{mnras}

\usepackage[english]{babel}
\usepackage{amsmath}
\usepackage{amssymb}
\usepackage{txfonts}
\usepackage{graphicx}
\usepackage{ulem}

\newcommand{\ud}{\ensuremath{\mathrm{d}}}

\title[Core-Collapse Supernovae from Low-Mass Progenitors]{Three-Dimensional Simulations of Neutrino-Driven Core-Collapse Supernovae from Low-Mass Single and Binary Star Progenitors}

\author[B.~M\"uller et al.]{
  Bernhard~M\"uller$^{1}$\thanks{E-mail: bernhard.mueller@monash.edu},
  Thomas~M.~Tauris$^{2}$,
  Alexander~Heger$^{1,3}$,
  Projjwal~Banerjee$^{4}$,\and
  Yong-Zhong~Qian$^{5,3}$,
  Jade~Powell$^{6}$,
  Conrad~Chan$^{1}$,
  Daniel~W.~Gay$^{7,1}$,
  Norbert~Langer$^{8,9}$
\\
$^1${Monash Centre for Astrophysics, School of
  Physics and Astronomy, Monash University, Victoria
  3800, Australia} \\
$^{2}$Department of Physics and Astronomy, Aarhus University, Ny Munkegade 120, 8000 Aarhus C, Denmark\\
$^{3}${Tsung-Dao Lee Institute, Shanghai 200240, China.} \\ 
$^{4}${Department of Astronomy, School of Physics and Astronomy, Shanghai Jiao Tong University, Shanghai 200240, China.}\\
$^{5}${School of Physics \& Astronomy,
  University of Minnesota, Minneapolis, MN 55455, U.S.A.} \\
$^{6}${OzGrav, Swinburne University of Technology, Hawthorn, VIC 3122, Australia} \\
$^{7}${Astrophysics Research Centre, School
  of Mathematics and Physics, Queen's University
  Belfast, Belfast, BT7~1NN, United Kingdom}\\
$^{8}${Argelander-Institut f\"ur Astronomie, Universit\"at Bonn, Auf dem H\"ugel 71, D-53121 Bonn, Germany}\\
$^{9}${Max-Planck-Institut für Radioastronomie, Auf dem H\"ugel 69, D-53121 Bonn, Germany}
}

\begin{document}

\label{firstpage}
\pagerange{\pageref{firstpage}--\pageref{lastpage}}

\maketitle

\begin{abstract}
We present a suite of seven 3D supernova simulations of
non-rotating low-mass progenitors using multi-group neutrino 
transport. Our simulations cover single star progenitors with 
zero-age main sequence masses between $9.6 M_\odot$ and 
$12.5 M_\odot$ and (ultra)stripped-envelope progenitors with initial
helium core masses between $2.8 M_\odot$ and $3.5 M_\odot$. We 
find explosion energies between $0.1\,\mathrm{Bethe}$ and 
$0.4\,\mathrm{Bethe}$, which are still rising by the end of the 
simulations. Although less energetic than typical events, our 
models are compatible with observations of less energetic 
explosions of low-mass progenitors. In six of our models, the 
mass outflow rate already exceeds the accretion rate onto the 
proto-neutron star, and the mass and angular momentum of the 
compact remnant have closely approached their final value,
barring the possibility of later fallback. While the
proto-neutron star is still accelerated by the gravitational tug
of the asymmetric ejecta, the acceleration can be extrapolated to 
obtain estimates for the final kick velocity. We obtain 
gravitational neutron star masses between $1.22 M_\odot$ and 
$1.44 M_\odot$, kick velocities between $11\, \mathrm{km}\, \mathrm{s}^{-1}$ and $695\, \mathrm{km}\, \mathrm{s}^{-1}$, and 
spin periods from $20\, \mathrm{ms}$ to $2.7\, \mathrm{s}$, which 
suggests that typical neutron star birth properties can be 
naturally obtained in the neutrino-driven paradigm. We find a 
loose correlation between the explosion energy and the kick 
velocity. There is no indication of spin-kick alignment, but a 
correlation between the kick velocity and the neutron star 
angular  momentum, which needs to be investigated further as a 
potential point of tension between models and observations.
\end{abstract}


\section{Introduction}
\label{sec:intro}
While it is well established that many massive stars end their life as
a core-collapse supernova, numerical simulations have long struggled
to conclusively explain the mechanism that powers these
explosions. The best-explored scenario is arguably the neutrino-driven
mechanism, which relies on the partial reabsorption of neutrinos
emitted from the young proto-neutron star (PNS) and the accretion
layer at its surface to revive the shock and power the explosion
\citep[for reviews, see][]{mezzacappa_05,janka_12,burrows_12}.  Models
of neutrino-driven explosions have long teetered on the verge between
success and failure, and temporary setbacks -- like the failure of the
first three-dimensional (3D) three-flavour multi-group neutrino
hydrodynamics simulations \citep{hanke_12} -- sometimes unduly
obscured the progress in methodology and understanding.

In the last few years, however, 3D first-principle models of
neutrino-driven explosions have become increasingly successful
\citep[see the reviews of][]{mueller_16b,janka_16}.  Shock revival by
neutrino heating has now been observed in a sizable number of 3D
simulations by different groups employing multi-group transport with
varying degrees of sophistication
\citep{takiwaki_12,takiwaki_14,melson_15a,melson_15b,mueller_15b,lentz_15,roberts_16,mueller_17,ott_18,chan_18,kuroda_18,summa_18,mueller_18,vartanyan_18}
for a wide mass range of progenitors. Perhaps even more importantly in
the light of modelling uncertainties and lingering failures without
shock revival \citep{hanke_12,tamborra_14a,oconnor_18b}, it has been
realized that a number of physical effects that are not yet included
in most simulations can systematically help to further expedite
neutrino-driven explosion, such as the softening of the equation of
state by muon creation \citep{bollig_17}, convective seed asymmetries
in the progenitors
\citep{couch_14,mueller_15a,couch_15,mueller_16b,mueller_17}, and the
reduction of the neutrino scattering opacity due to nucleon
correlations \citep{horowitz_17,bollig_17}, or the strangeness of the
nucleon \citep{melson_15b}.

With strong indications that neutrino heating can indeed trigger shock
revival, one of the next challenges for the models is to predict
explosion and compact remnant properties in order to connect to
observational findings, e.g., on the distribution and systematics of
neutron star birth masses \citep{oezel_16,antoniadis_16,tauris_17}, neutron star kicks \citep{hobbs_05,faucher_06,ng_07},
and supernova explosion energies and nickel masses
\citep{poznanski_13,pejcha_15c,mueller_t_17}. First-principle simulations still face a time-scale
problem here: Whereas shock revival is typically expected to
occur on time scales of only a few hundred milliseconds after the
collapse of the iron core to a PNS, the explosion and remnant
properties are determined during a phase of concurrent accretion and
mass ejection that can last on the order of seconds
\citep{mueller_15b,bruenn_16}. While axisymmetric models can probe
these long time scales \citep{mueller_15b,bruenn_16,nakamura_16}, 3D
effects qualitatively alter this phase \citep{mueller_15b}. Except for
low-energy explosions from the least massive progenitors
\citep{melson_15a,mueller_18}, self-consistent 3D models cannot yet
follow neutrino-driven explosions sufficiently far to obtain fully
converged explosion energies and remnant properties. It thus remains
to be demonstrated that neutrino-driven explosions can explain the
full gamut of supernova energies -- although the long-time models of
\citet{bruenn_16} in two dimensions (2D) and \citet{mueller_17} in 3D
have progressed far towards this goal -- and reveal the systematics of
explosion and remnant properties. More phenomenological models
\citep{ugliano_12,ertl_15,pejcha_15a,mueller_16a,sukhbold_16} are
presently the only viable theoretical approach for this purpose, but
it is imperative that they be put on a firmer footing with the help of
select multi-dimensional simulations.

In this paper, we take the next step towards this goal. Using a larger
set of long-time 3D simulations with multi-group transport, we explore
variations in explosion and remnant properties of supernova
progenitors with low-mass cores and investigate possible correlations
among them, such as the claim of spin-kick alignment \citep{ng_07,noutsos_12,noutsos_13,rankin_15} and
the suggested correlation between explosion energy, progenitor mass,
and neutron star kick
\citep{bray_16,janka_17,tauris_17,vigna_gomez_18}. Focusing on progenitors with
low-mass cores allows us to advance the simulations sufficiently far
to tentatively address these questions using 3D models with
multi-group neutrino transport for the first time. Recognizing the
importance of binary star evolution for stripped-envelope supernovae
\citep{smith_11,eldridge_13}, we include both single-star progenitors in the mass range
of  $9\text{-}13 M_\odot$ as well as binary progenitor
models, extending our recent work on ultra-stripped supernovae
\citep{mueller_18}. The single-star and binary-star progenitor models
cover a similar range of final helium core\footnote{Strictly speaking, the binary progenitors no longer have a helium core at collapse, as the helium shell forms the (tiny) envelope of the star. In case of the binary models, the ``final helium core mass'' is to be understood as the final mass that is left of the former helium star that was subsequently stripped further by Case~BB mass transfer.}
masses below $3.2M_\odot$.
The existence of ultra-stripped supernovae was first suggested purely based on
theoretical calculations of the final stages of mass transfer in close-orbit binaries \citep{tauris_13} 
and it was realized that progenitors of neutron star gravitational wave mergers must have
experienced such ultra-stripped supernovae when the second-formed neutron star is born \citep{tauris_13,tauris_15}.
A number of promising ultra-stripped supernova candidates have been identified based on
observed properties of rapidly decaying supernova light curves
 \citep{drout_13,moriya_17,de_18}.

Some of the progenitors investigated here are initialized with
proper 3D initial conditions in the active oxygen burning shell by
simulating the last minutes of the shell evolution in 3D in the vein
of \citet{mueller_16c}.

Our paper is structured as follows: In Section~\ref{sec:progenitors}
we introduce the stellar progenitor models, including a short
discussion of the multi-dimensional flow geometry in the oxygen shell
at collapse where applicable.  In Section~\ref{sec:numerics}, we
briefly describe the \textsc{CoCoNuT-FMT} code used for our supernova
simulations with a focus on recent updates to the neutrino rates and
hydrodynamics. The results of the simulations are presented in
Section~\ref{sec:results}, and we conclude with a discussion of their
implications in Section~\ref{sec:conclusions}.

\begin{table*}
    \centering
        \caption{Summary of initial stellar models and simulation setup.}
    \label{tab:setup}
    \begin{tabular}{ccrrcccccccccc}
    \hline
              & Initial  & $M_\mathrm{ini}$ & $M_\mathrm{f}$ & $M_\mathrm{He,f}$&Z & Code up to      & 3D initial & PPM & axis  & nucleon & weak \\
        Model & state    & $[M_\odot]$   & $[M_\odot]$     & $[M_\odot]$     & $[Z_\odot]$ & Ne burning & conditions & scheme & treatment & correlations & magnetism & $g_\mathrm{a,s}$\\
    \hline
         z9.6  & H ZAMS & 9.6 & 9.6 & 1.38 & 0     & \textsc{Kepler} & no  & CW84 & coarsening & no  & no & 0\\
         s11.8 & H ZAMS & 11.8 & 10.4 & 2.45 &1 & \textsc{Kepler} & yes & CS08 & coarsening & yes  & no & -0.05\\
         z12   & H ZAMS & 12.0 & 12.0  & 2.49 & 0     & \textsc{Kepler} & yes & CW84 & coarsening & yes & no & -0.05\\
         s12.5 & H ZAMS & 12.5 & 9.78 & 3.17 & 1 & \textsc{Kepler} & yes & CW84 & coarsening & yes  & no & 0\\
         he2.8 & He ZAMS & 2.8 & 1.49 & 1.49& 1 & \textsc{BEC}    & no  & CW84 & coarsening & no  & no & 0\\
         he3.0 & He ZAMS & 3.0  & 3.00 & 3.00 & 1  & \textsc{Kepler} & yes  & CS08 & Fourier filter& yes  & yes & -0.05\\
         he3.5 & He ZAMS & 3.5  & 2.39&  2.39 &1  &  \textsc{BEC}   & no  & CS08 & Fourier filter& yes  & no & -0.05\\
          \hline
    \end{tabular}
    \flushleft
    $M_\mathrm{ini}$ is the initial mass of the stellar evolution model. For the single star progenitors, this is the zero-age main sequence mass;
for the helium star (binary) models, it is the initial helium star mass.
$M_\mathrm{f}$ and  $M_\mathrm{He,f}$ are the progenitor mass 
 and the helium (core) mass at the pre-supernova stage; these are identical for the stripped-envelope models.
$Z$ is the initial metallicity. 
CW84 refers to the original PPM reconstruction of \citet{colella_84},
and CS08 refers to the 6th-order extremum-preserving
method of \citet{colella_08}. $g_\mathrm{a,s}$ is the
strangeness contribution to the axial coupling constant for neutral currents.
\citet{tauris_15}.
\end{table*}

\begin{figure}
\includegraphics[width=\linewidth]{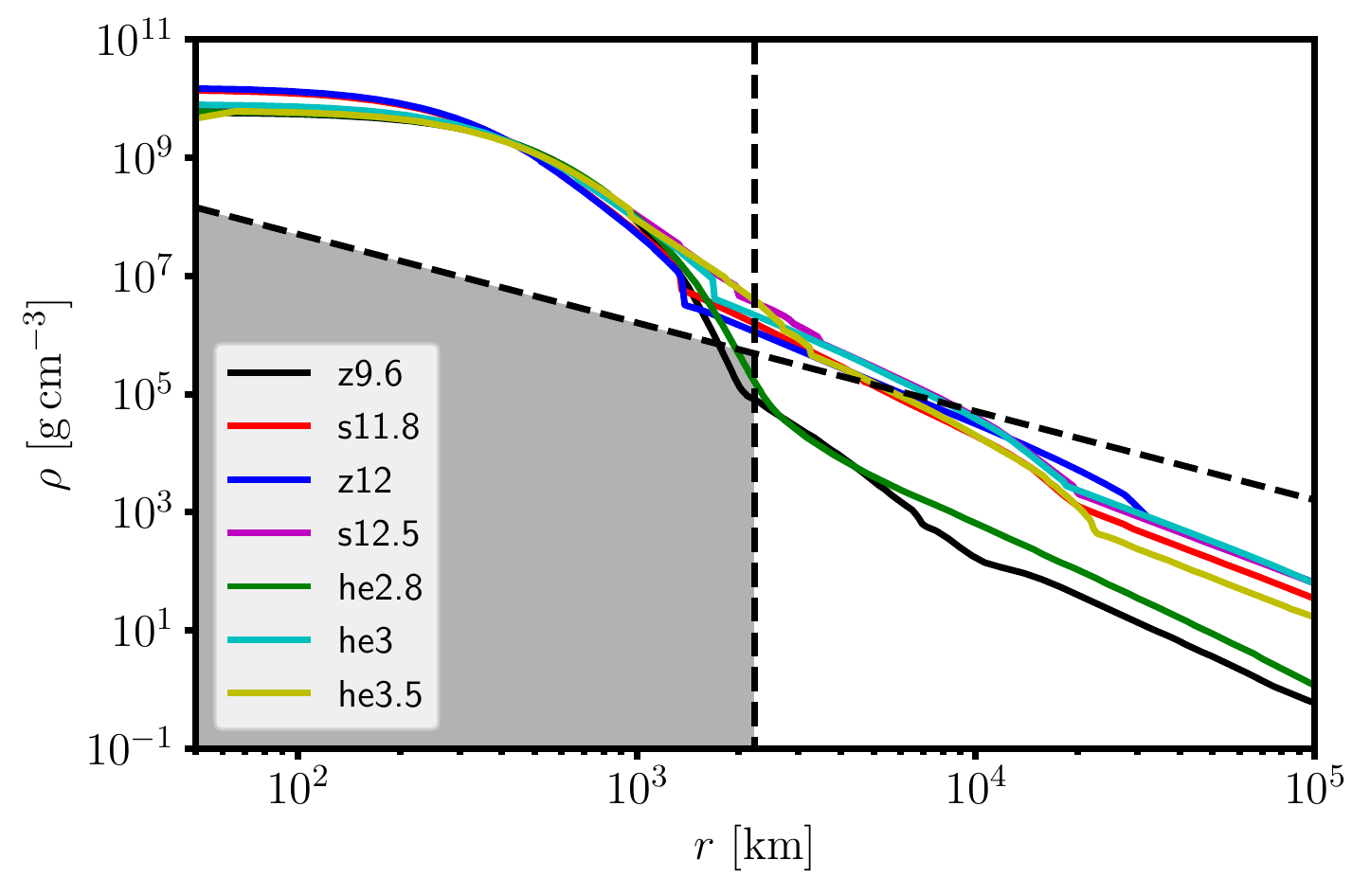}
\caption{Density profiles of the seven progenitor models. The
grey-shaded area marks the region for ECSN-like explosion
dynamics, which obtains if the density drops below
a critical value given by Equation~(\ref{eq:rhocrit})
(slanted dashed line) inside a radius of about $2200 \, \mathrm{km}$ (vertical dashed line). The slanted dashed line also roughly corresponds
to an accretion rate onto the shock of $0.05 M_\odot \, \mathrm{s}^{-1}$.
Models z9.6 and he2.8 fall into the regime of ECSN-like explosions.
\label{fig:profiles}}
\end{figure}

\begin{figure*}
\includegraphics[width=0.48 \textwidth]{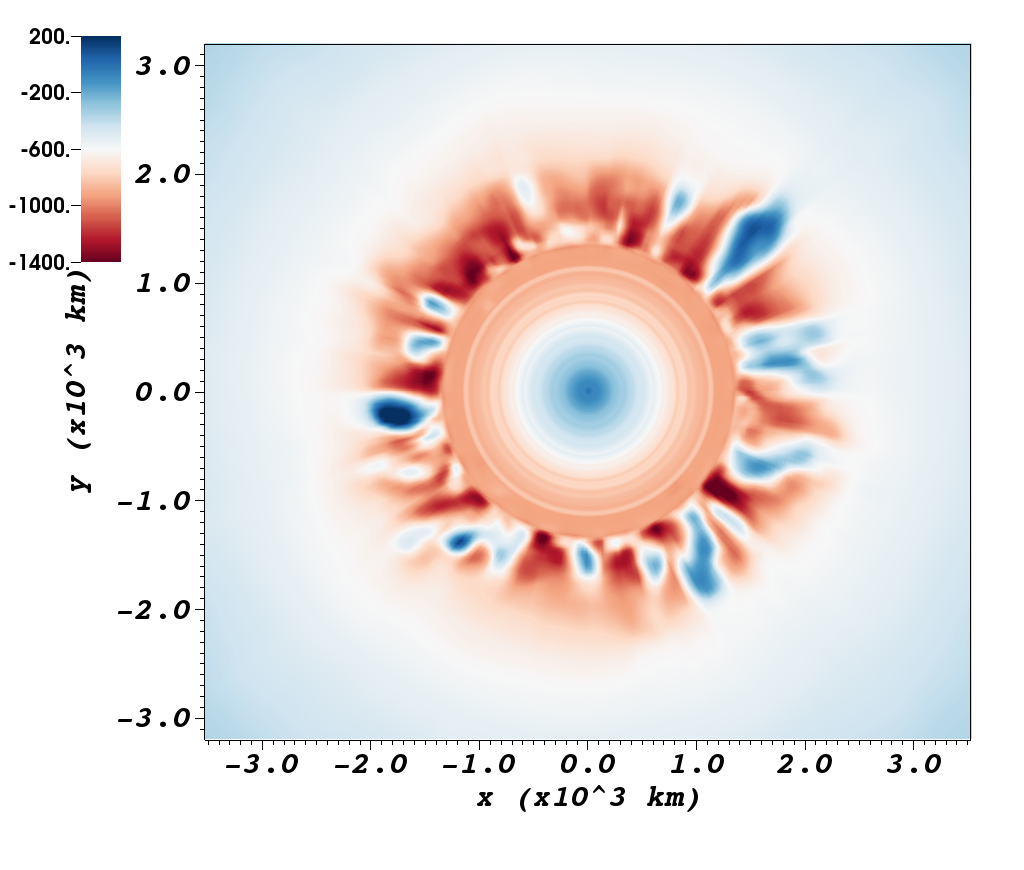}
\includegraphics[width=0.48 \textwidth]{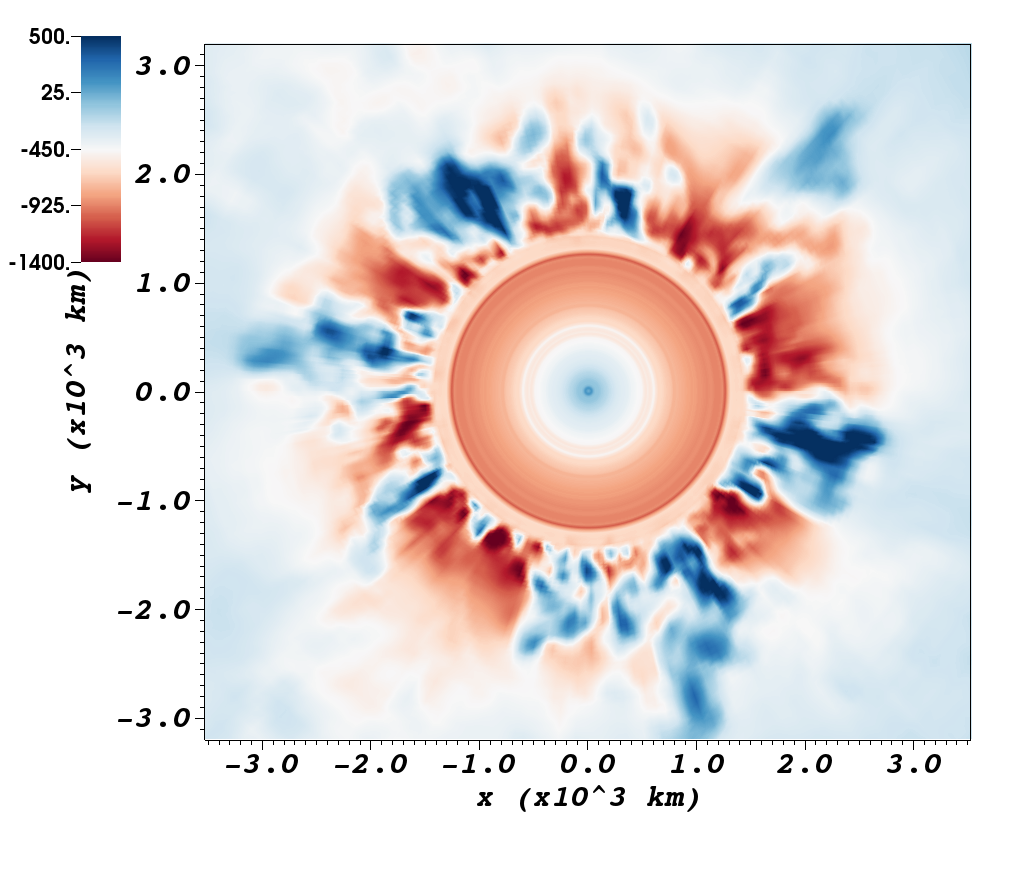}\\
\includegraphics[width=0.48 \textwidth]{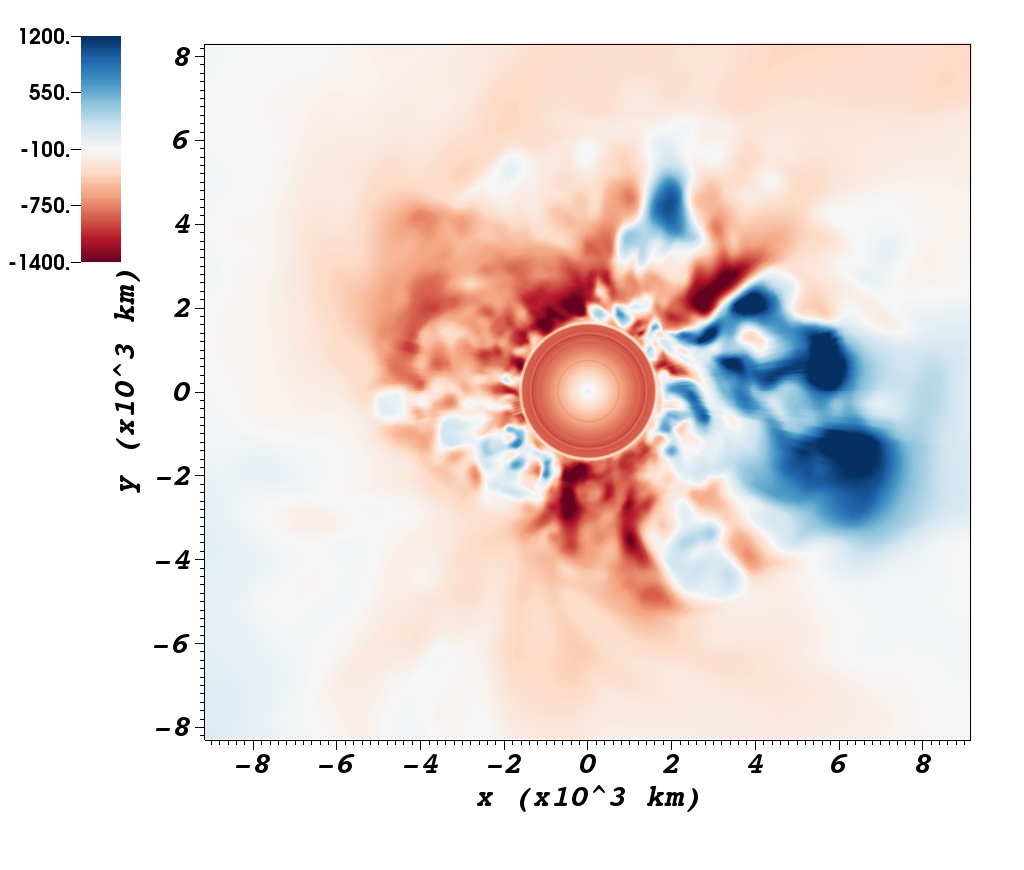}
\includegraphics[width=0.48 \textwidth]{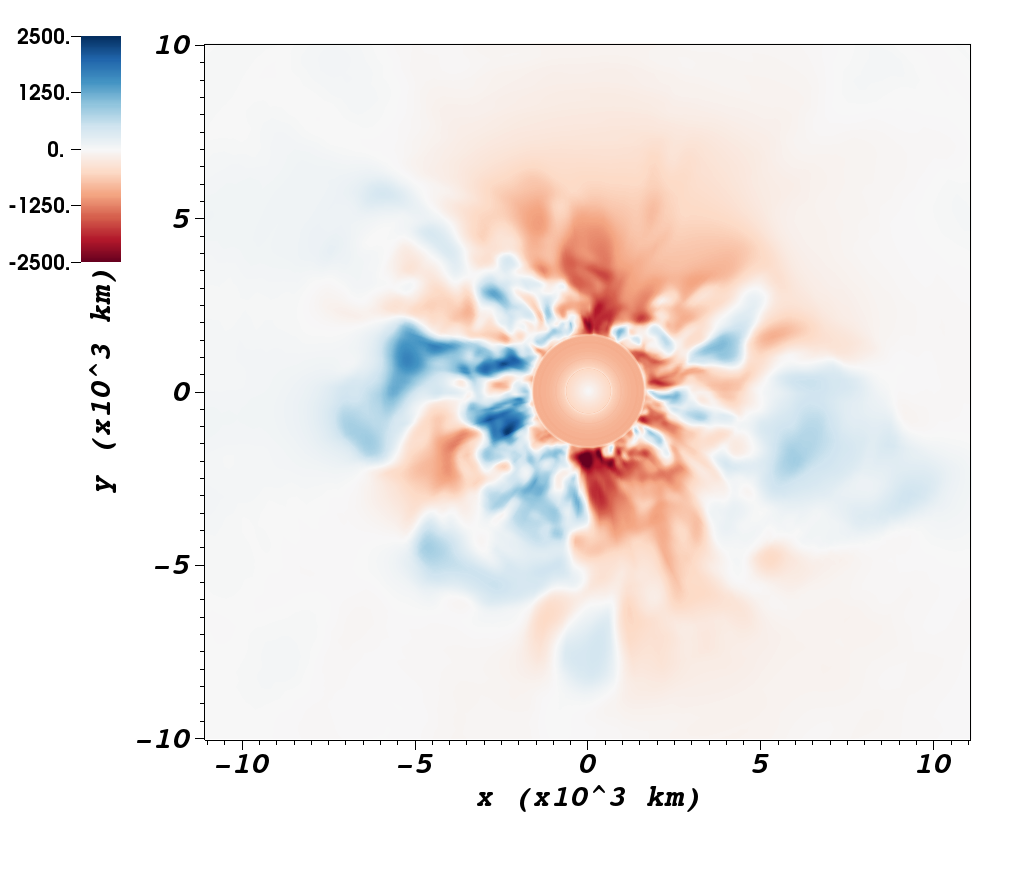}
\caption{Radial velocity 
in units of $100 \, \mathrm{km}\, \mathrm{s}^{-1}$ for progenitor
models s11.8 (top left), z12 (top right),
s12.5 (bottom left), and he3.0 (bottom right) on 2D slices at the onset
of collapse. The models exhibit widely
varying convective velocities
and flow geometries in the active oxygen shell.
\label{fig:prog3d}}
\end{figure*}

\section{Progenitor Models}
\label{sec:progenitors}
We simulate the collapse and explosion of  seven progenitor
models of different mass and metallicity. Four of our progenitors
(z9.6, s11.8, z12, and s12.5) are single star models evolved from the
hydrogen zero-age main sequence (ZAMS) to core collapse using the
stellar evolution code \textsc{Kepler} \citep{weaver_78,heger_10}.  Two
of these have been described previously: Model z9.6 was first used in
\citet{janka_12b}, and the 3D simulation presented here is identical
to the one in \citet{mueller_16b}; progenitor model s11.8 is taken from
\citet{banerjee_16}.

In addition to single star progenitors, we consider progenitors that
have undergone mass loss via binary interactions. Models he2.8 and
he3.5 \citep{tauris_15} are examples for ultra-stripped progenitors
that can evolve from the companion star in high-mass X-ray binary (HMXB) 
systems which undergo common-envelope (CE) evolution followed by an additional
(so-called Case~BB) mass-transfer episode \citep{tauris_13,tauris_15,tauris_17}. Model he2.8 is the
same progenitor as used in \citet{mueller_18} with an initial helium core
mass of $2.8 M_\odot$ and an initial orbital period of $20 \,
\mathrm{d}$. Model he3.5 has an initial helium core mass of $3.5
M_\odot$ and an initial orbital period of $2 \, \mathrm{d}$, and is
representative for the ultra-stripped progenitors with more massive
metal cores at collapse in \citet{tauris_15}. These two models were
evolved starting with a zero-age helium star main-sequence configuration after the termination of
the CE phase. The interior evolution and the final mass transfer were
calculated using the binary evolution code \textsc{BEC}
\citep{wellstein_01,yoon_10} until after the detachment of the binary. 
To follow the late burning stages using a large nuclear reaction network,
the models were then mapped to \textsc{Kepler} during core
Ne burning.

In addition, we consider a $3 M_\odot$ helium star evolved with
\textsc{Kepler} as an example of a progenitor that has lost its
complete H envelope due to binary interactions, but has not undergone
further mass transfer afterwards.

Both the hydrogen-rich and (ultra)stripped-envelope progenitors cover
a similar range in final helium core mass $M_\mathrm{He,f}$, reaching
as low as $M_\mathrm{he,f}=1.38 \, M_\odot$ in case of z9.6 and as
high as $M_\mathrm{he,f}=3.17 \, M_\odot$ in case of s12.5
(see Table~\ref{tab:setup}). Due to their small helium core
mass, the two most extreme models z9.6 and he2.8 exhibit
 a steep density gradient at the edge of their Si core as can
be seen from their density profiles in Figure~\ref{fig:profiles}. 
Although the density gradient is not as steep as for electron-capture
supernova (ECSN) progenitors, models z9.6 and he2.8 still fall in
the regime of ``ECSN-like'' progenitors, which are characterised
by a rapid drop of the mass accretion rate between a critical
value of $\dot{M}_\mathrm{crit}\sim 0.05 M_\odot \, \mathrm{s}^{-1}$ within the
first few hundred milliseconds after core bounce.
This  requires that
the density $\rho$ drops below
\begin{equation}
\label{eq:rhocrit}
    \rho<\frac{1}{8}\sqrt{\frac{3}{Gm}}\dot{M}_\mathrm{crit} r^{-3/2},
\end{equation}
in terms of the mass coordinate $m$ and radius $r$ inside
a radius of $\mathord{\approx} 2200 \, \mathrm{km}$
\citep{mueller_16b}. The other models (s11.8, z12, s12.5, he3, he3.5)
remain well above this threshold.

Recognising the potentially significant role of pre-collapse seed
asphericities in the core-collapse supernova explosion mechanism
\citep{couch_13,mueller_15a,mueller_17}, we simulated the convective
burning in the active O shell in 3D for the last few minutes prior to
collapse for some of these progenitors (s11.8, z12, s12.5, and he3.0)
using the same methodology as in \citet{mueller_16c}.  In
Figure~\ref{fig:prog3d} we show the radial velocity at the onset of
core collapse on 2D slices for these models.  Our 3D simulations of O
shell burning confirm the predictions of \citet{collins_18} inasmuch
as they show significant variations in convective velocities, Mach
numbers, and flow geometries across progenitors.  Model s11.8 exhibits
comparatively weak small-scale convection in the O shell with a
convective Mach number of 0.06 at the base of the shell, Model z12 is
relatively similar to the $18 M_\odot$ progenitor of
\citet{mueller_16c} in terms of the convective Mach number, but the
typical convective eddy scale is comparatively small. Models s12.5
and he3 both have thick convective shells dominated by large-scale
motions and even higher convective Mach numbers of 0.14 and 0.16 at
the base.

Some key parameters of the progenitor models, as well as
details of the core-collapse supernova simulations that
will be described in the next section, are summarised
in Table~\ref{tab:setup}.

\section{Numerical Methods and Simulation Setup}
\label{sec:numerics}

We use the \textsc{CoCoNuT-FMT} code, which solves the equations of general
relativistic neutrino hydrodynamics using the extended conformal
flatness (xCFC) approximation \citep{cordero_09} for the space-time
metric.  The finite-volume hydro solver employs a hybrid HLLC/HLLE
Riemann solver with higher order reconstruction
\citep{dimmelmeier_02_a,mueller_10} and uses spherical polar
coordinates. As in \citet{mueller_15b}, severe time-step constraints
in 3D are avoided by treating the core of the PNS in 1D and adopting a
mixing-length treatment for PNS convection, and by using a filtering
scheme for the conserved variables near the axis.

Recent updates to the code include 6th-order extremum preserving
reconstruction \citep{colella_08,sekora_09} instead of the original
4th-order piecewise parabolic method of \citet{colella_84} and an
alternative scheme for filtering near the polar axis.
\citet{mueller_15b} and follow-up work relied on a mesh coarsening
scheme that involved averaging the solution on the original spherical
polar grid over coarser ``supercells'' and projecting back to the fine
grid using piecewise linear reconstruction, which may favour
the development of bipolar asymmetries in weakly perturbed models. For
the more recent models, we have therefore adopted an alternative
scheme for taming the prohibitive Courant constraint near the grid
axis. Following common practice in meteorology \citep{jablonowski_11}, we apply a
latitude-dependent filter to damp short wavelength Fourier modes. This
technique avoids the axis artefacts of the older mesh coarsening
technique and allows us to smoothly switch off filtering at large
radii without complicating the the data layout and communication in
MPI mode.  The implementation of both the mesh coarsening scheme and
the Fourier-based filter is described in more detail in Appendix~\ref{sec:filter}.

For the neutrinos, we use the fast multi-group transport scheme of
\citet{mueller_15a}, which solves the frequency-dependent neutrino
energy equation assuming stationarity and a one-moment closure obtained
from the solution of a simplified Boltzmann equation in the two-stream
approximation and an analytic closure at low optical depth. The scheme
accounts for gravitational redshift, but largely ignores
velocity-dependent terms.  Compared to the original implementation of
\citet{mueller_15a}, we do however include a Doppler correction term for the
absorption opacity $\kappa_\mathrm{a}$ in the vein of a mixed-frame
formulation \citep[cp.][]{hubeny_07} when we solve the zeroth moment
equation after the flux factor has been determined.  Using the fact
that $\kappa_\mathrm{a}$ is roughly proportional to the square of the
neutrino energy to avoid the numerical calculation of opacity
derivatives, we modify it according to
 \begin{equation}
     \kappa_\mathrm{a} \rightarrow \kappa_\mathrm{a}
     \left[W (1-v_r f_H/c)\right]^2,
 \end{equation}
in terms of the radial velocity $v_r$, the Lorentz
factor $W=(1-v^2)^{1/2}$, and the flux factor $f_H$.
 
All our models include the effect of nucleon interaction potentials on
the charged-current rates \citep{martinez_12}; and in the more recent
simulations, we include the modification of the neutrino-nucleon
scattering opacity due to nucleon correlations following
\citet{horowitz_17}. Model he3.0 also includes weak magnetism
corrections following \citet{horowitz_02}.

At high densities, we use the nuclear equation of state of
\citet{lattimer_91} with a bulk incompressibility of $K=220 \,
\mathrm{MeV}$, supplemented by a low-density equation of state for
nuclei, nucleons, and lepton and photon radiation. The flashing
treatment of \citet{rampp_02} is used for nuclear reactions below a
temperature of $0.5\, \mathrm{MeV}$; at higher temperatures, nuclear
statistical equilibrium is assumed.

\begin{figure}
    \centering
    \includegraphics[width=\linewidth]{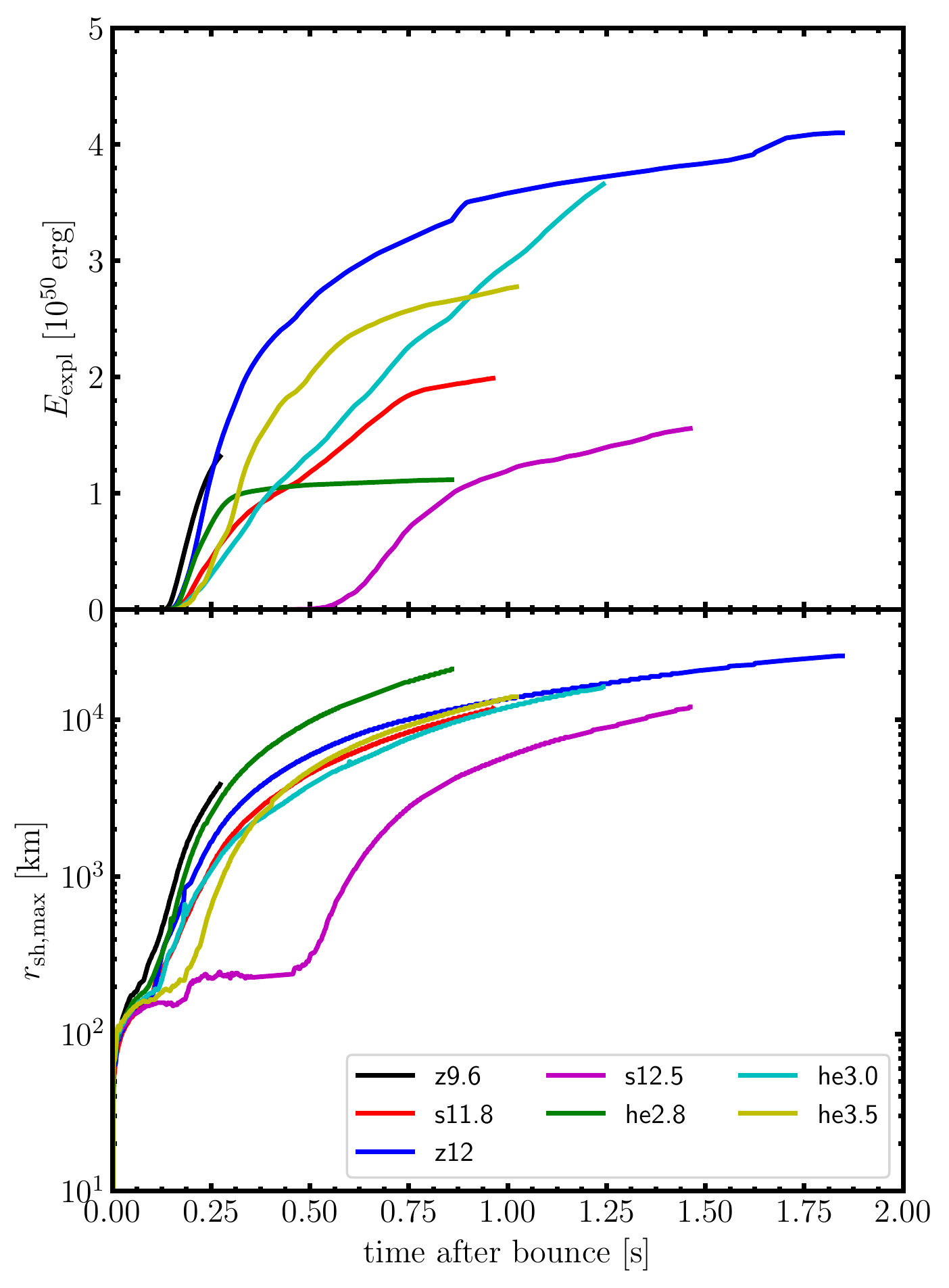}
    \caption{Diagnostic explosion energy $E_\mathrm{diag}$ (top) 
    and maximum shock radius $r_\mathrm{sh,max}$ as a function of time
    for all of the seven low-mass models.
    \label{fig:energy}}
\end{figure}

\begin{figure*}
\includegraphics[width=0.48 \textwidth,trim={1cm 1cm 1cm 1cm},clip]{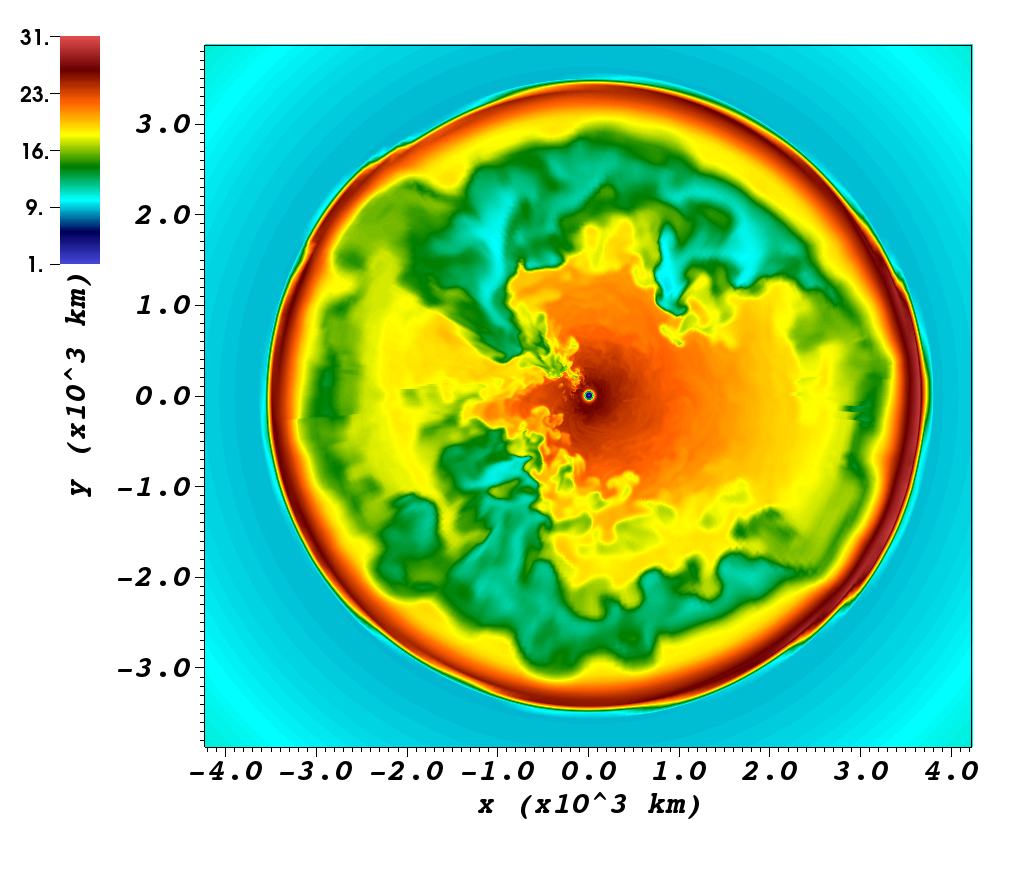}
\includegraphics[width=0.48 \textwidth,trim={0.8cm 1.3cm 0.4cm 0cm},clip]{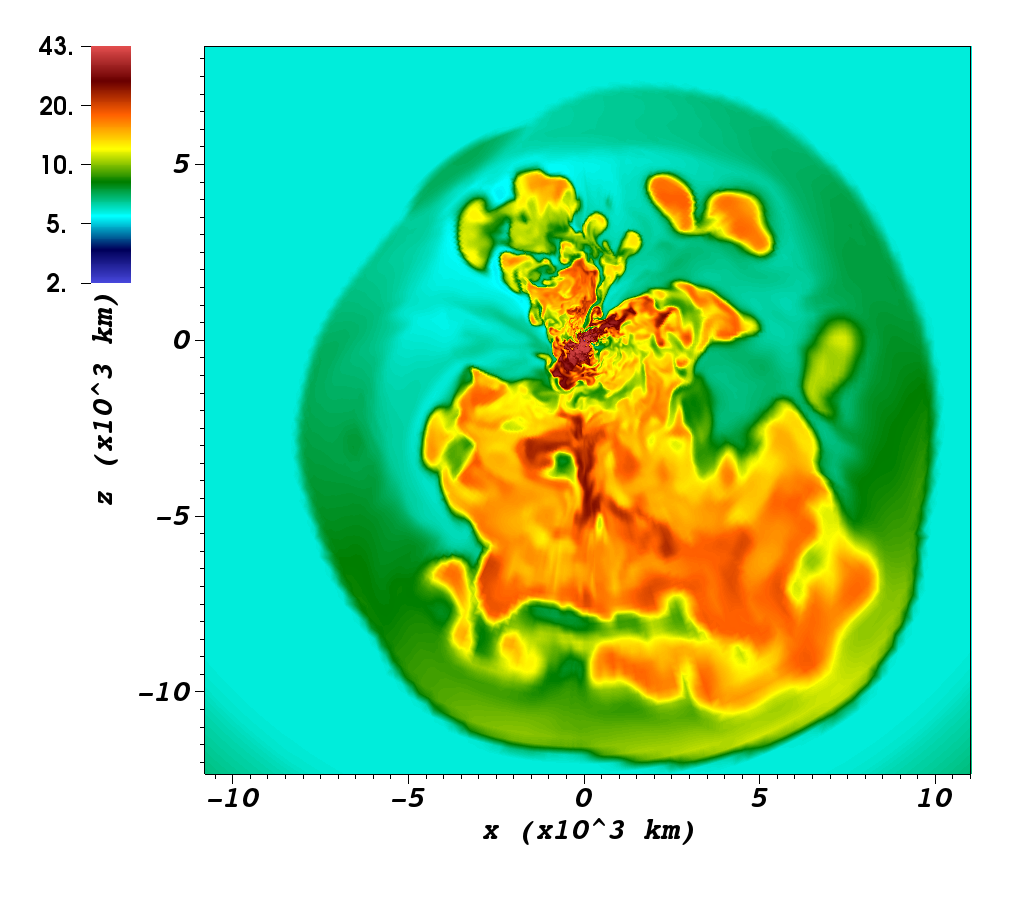}\\
\includegraphics[width=0.48 \textwidth,trim={1cm 1cm 1cm 1cm},clip]{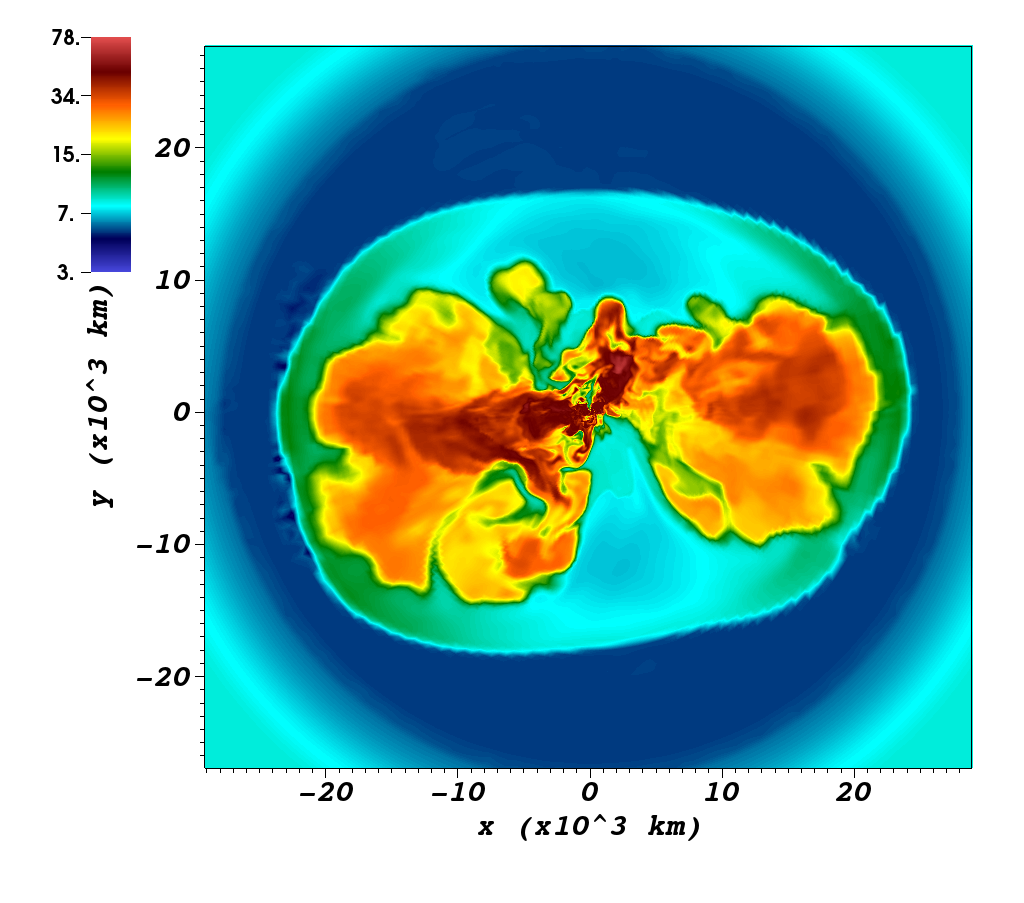}
\includegraphics[width=0.48 \textwidth,trim={1cm 1cm 1cm 1cm},clip]{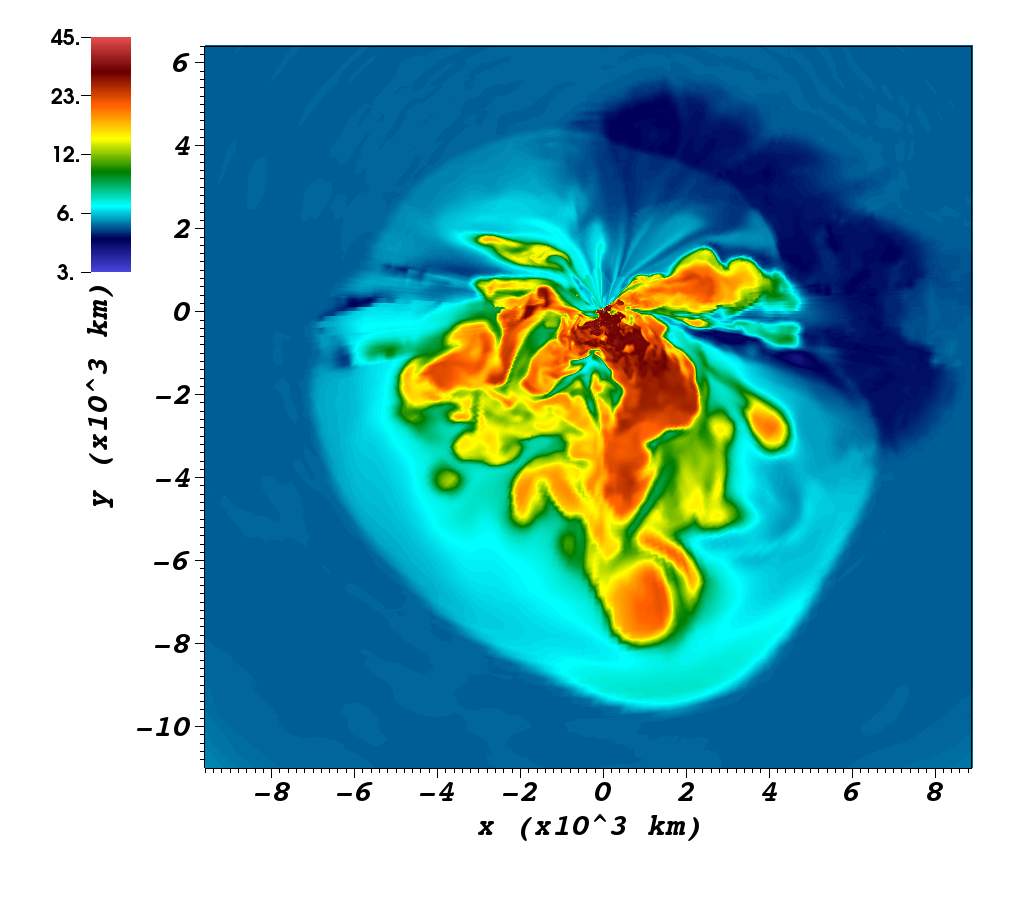}\\
\includegraphics[width=0.48 \textwidth,trim={1cm 1cm 1cm 1cm},clip]{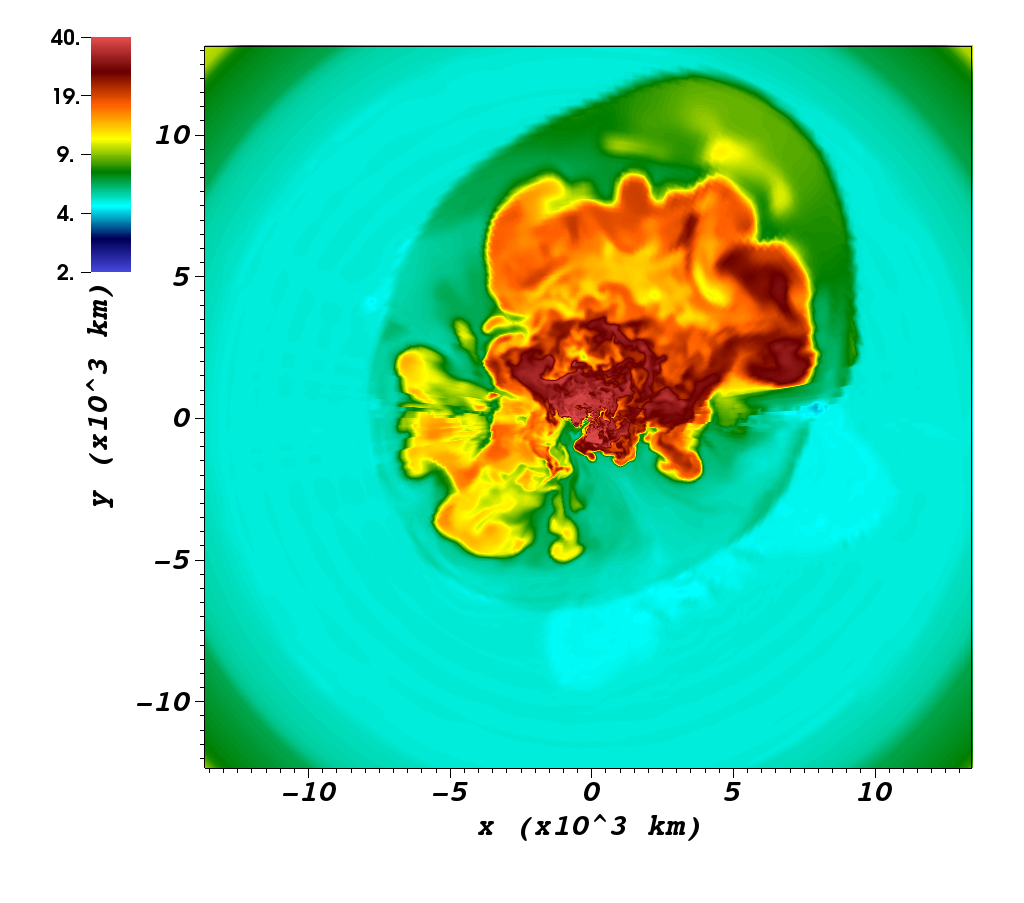}
\includegraphics[width=0.48 \textwidth,trim={1cm 1cm 1cm 1cm},clip]{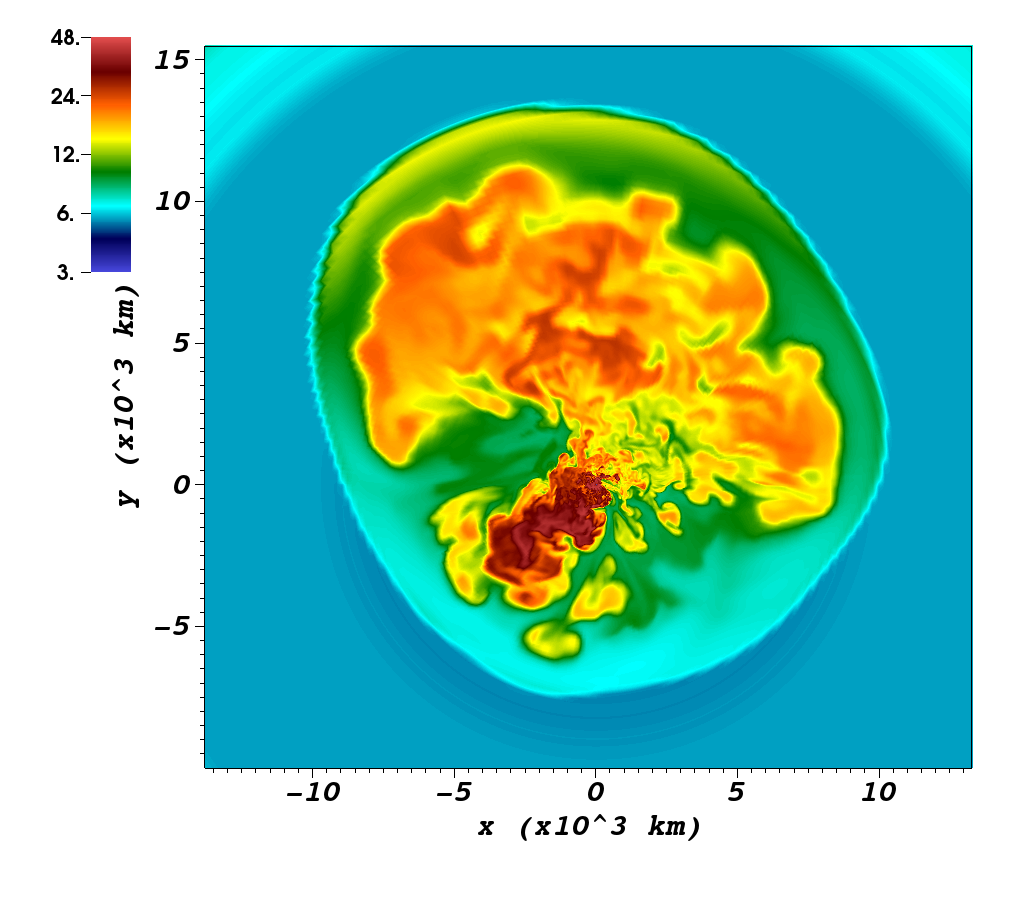}\\
\caption{Entropy $s$ in units of $k_\mathrm{b}/\mathrm{nucleon}$
on 2D slices at the end of the simulations for models z9.6 (top left), 
s11.8 (top right), z12 (middle left), s12.5 (middle right), he3 (bottom 
left) and he3.5 (bottom right). The axis of the spherical polar grid is
aligned with the $x$-axis of the plots. Note that there is no visible
alignment of the flow structures with the axis of the spherical polar grid in models
s11.8, s12.5, he3, and he3.5. The explosion are predominantly unipolar,
with the exception of z12, and to some degree z9.6 at early times.
\label{fig:slices}}
\end{figure*}

\begin{table*}
    \centering
    \caption{Explosion and neutron star properties
    \label{tab:expl_prop}}
    \begin{tabular}{crccccrrrc}
    \hline
               & $t_\mathrm{fin}$ & $E_\mathrm{expl}$ & $M_\mathrm{IG}$ & $M_\mathrm{by}$ & $M_\mathrm{grav}$ & $v_\mathrm{PNS}$ 
               & $v_\mathrm{PNS,ex}$  & $P_\mathrm{PNS}$ & $\alpha$\\ 
         Model & (ms) & $(10^{50} \, \mathrm{erg})$   & $(M_\odot)$       &         $(M_\odot)$       & $(\mathrm{km}\, \mathrm{s}^{-1})$ & $(\mathrm{km}\, \mathrm{s}^{-1})$ &            $(\mathrm{ms})$\\
         \hline
         z9.6 & 273   & 1.32 & 0.014 & 1.35 & 1.22 & 9.2 & 21 & 1060 & $48^\circ$\\
         s11.8 & 963  & 1.99 & 0.024 & 1.35 & 1.23 & 164 & 278 & 152 & $64^\circ$\\
         z12   & 1847 & 4.10 & 0.039 & 1.35 & 1.22 & 58  & 64 & 205 & $62^\circ$\\
         s12.5 & 1461 & 1.56 & 0.013 & 1.61 & 1.44 & 170 & $>170$& 20 & $55^\circ$\\
         he2.8 & 860  & 1.12 & 0.010 & 1.42 & 1.28 & 10.4& 11 & 2749 & $55^\circ$\\
         he3.0 & 1242 & 3.66 & 0.035 & 1.48 & 1.33 & 308 & 695 & 93 & $76^\circ$\\
         he3.5 & 1023 & 2.78 & 0.031 & 1.57 & 1.41 & 159 & 238 & 98 & $80^\circ$\\
         \hline
    \end{tabular}
    
    $t_\mathrm{fin}$ is the final post-bounce time reached by each 
    simulation, $E_\mathrm{expl}$ is the final diagnostic explosion 
    energy at the end of the simulations, $M_\mathrm{IG}$ is the mass of
    iron-group ejecta, $M_\mathrm{grav}$ is the gravitational neutron star mass, $v_\mathrm{PNS}$ is the kick velocity at the end of the run, $v_\mathrm{PNS,ex}$ is the extrapolated kick obtained
    from Equation~(\ref{eq:extrapolation}), $P_\mathrm{PNS}$ is the
    estimated neutron star spin period, and $\alpha$ is the angle
    between the spin and kick vector at the end of the simulations.
\end{table*}

\section{Results}
\label{sec:results}
\subsection{Shock Propagation and Explosion Energetics}
\begin{figure}
    \centering
    \includegraphics[width=\linewidth]{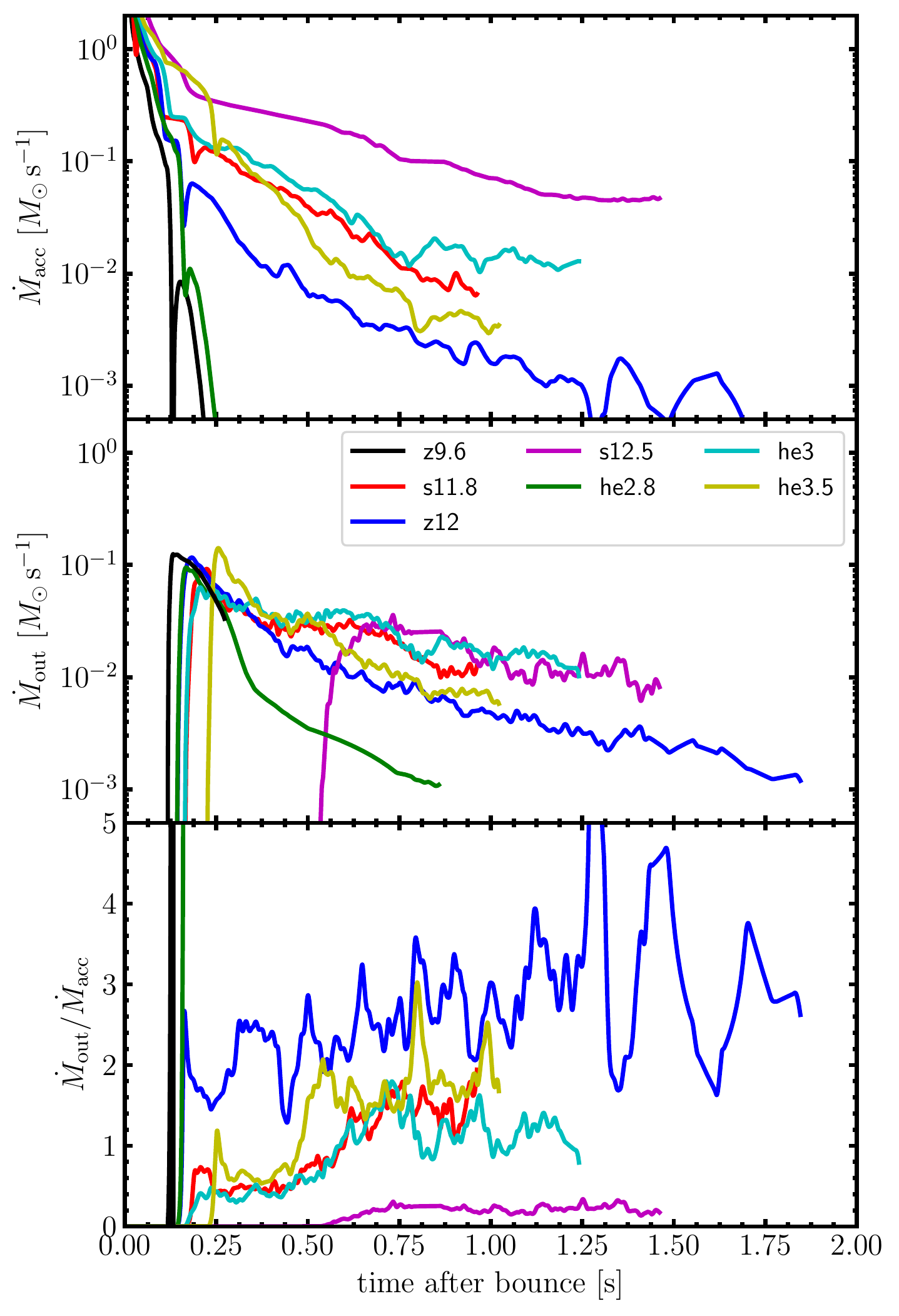}
    \caption{Mass accretion rate $\dot{M}_\mathrm{acc}$ (top),
    mass outflow rate $\dot{M}_\mathrm{out}$ (middle), and their
    ratio $\dot{M}_\mathrm{out}/\dot{M}_\mathrm{acc}$ for all models,
    measured at a radius of $400 \, \mathrm{km}$.
    \label{fig:inflow_outflow}}
\end{figure}

The evolution of the maximum shock radius and the diagnostic
explosion energy $E_\mathrm{expl}$ (defined as in
\citealt{mueller_17}) is shown in Figure~\ref{fig:energy}.  All the
models in this study undergo neutrino-driven shock revival and evolve
into explosion with large-scale unipolar or bipolar asymmetries
(Figure~\ref{fig:slices}). Important explosion and remnant
properties are summarised in Table~\ref{tab:expl_prop}.
With the exception of model s12.5, the shock is revived at rather early
post-bounce times between $100 \, \mathrm{ms}$ and $200 \,
\mathrm{ms}$. This is the result of the early infall of the O shell and
the concomitant drop of the mass accretion rate $\dot{M}_\mathrm{acc}$
(top panel of Figure~\ref{fig:inflow_outflow}).

\begin{figure}
    \centering
    \includegraphics[width=\linewidth]{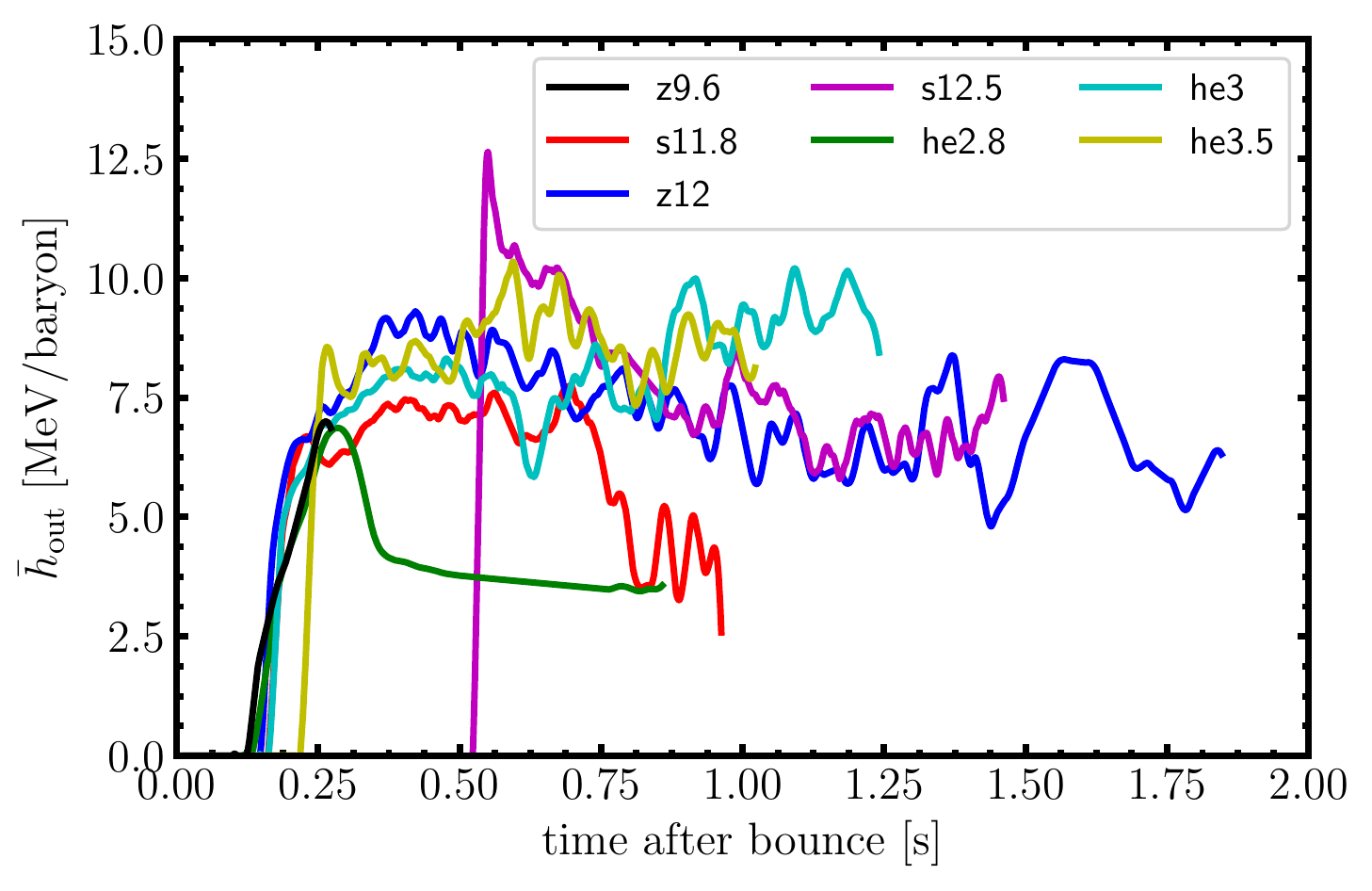}
    \caption{Specific total enthalpy $\bar{h}_\mathrm{out}$ (i.e.\ enthalpy minus gravitational binding energy) in the outflows as a function of post-bounce time for all models.
    \label{fig:enthalpy}}
\end{figure}

\begin{figure*}
\includegraphics[width=0.48 \textwidth,trim={0.5cm 1cm 1cm 1cm},clip]{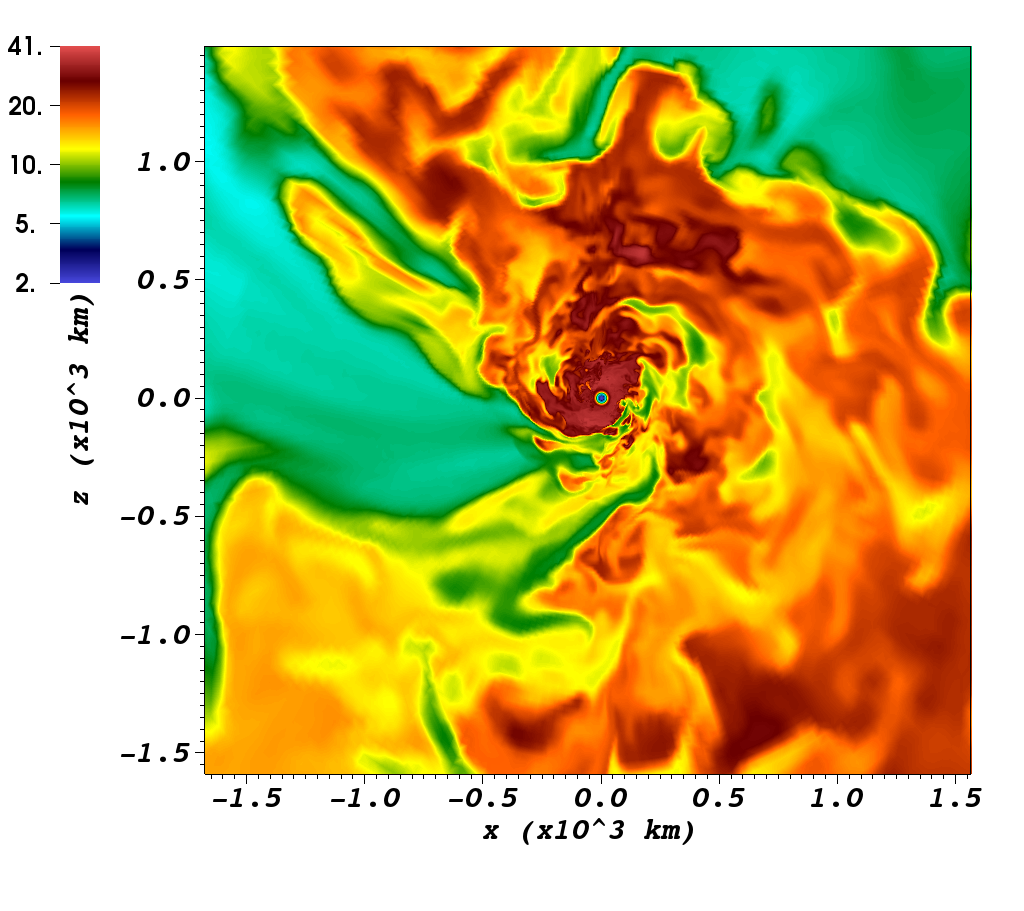}
\includegraphics[width=0.48 \textwidth,trim={0.5cm 1cm 1cm 1cm},clip]{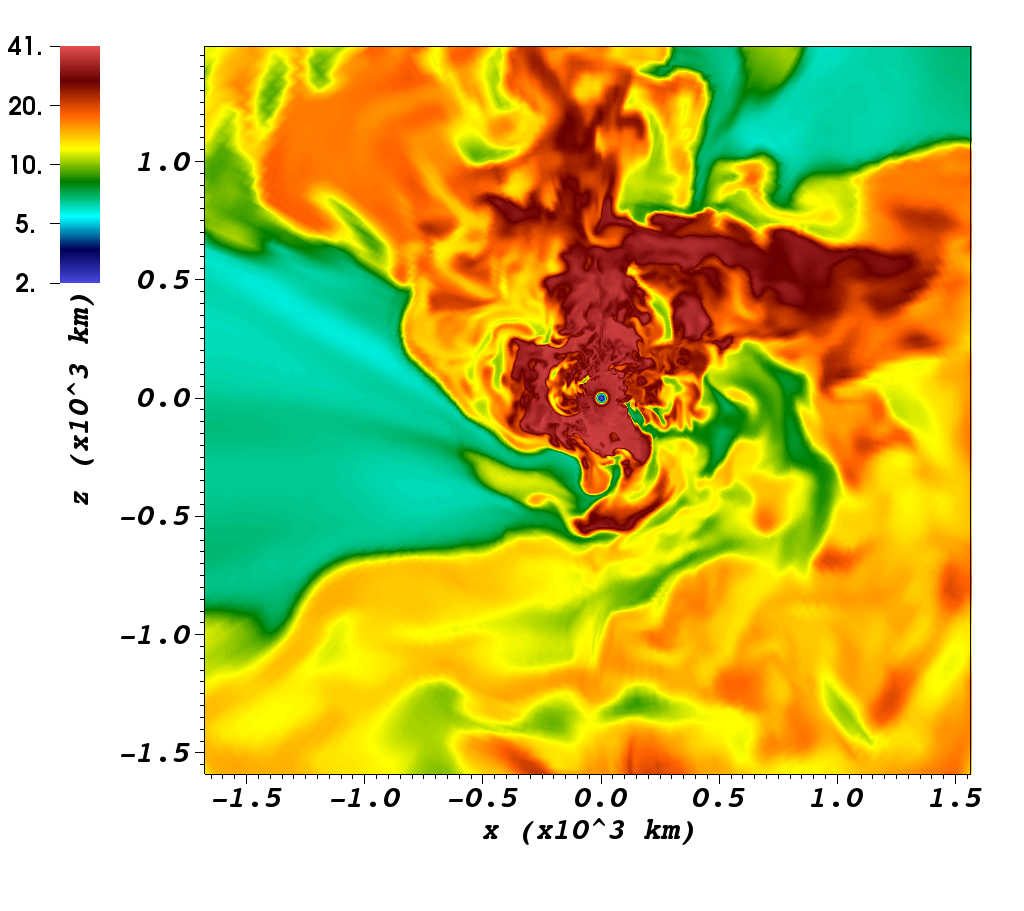}
\includegraphics[width=0.48 \textwidth,trim={0.5cm 1cm 1cm 1cm},clip]{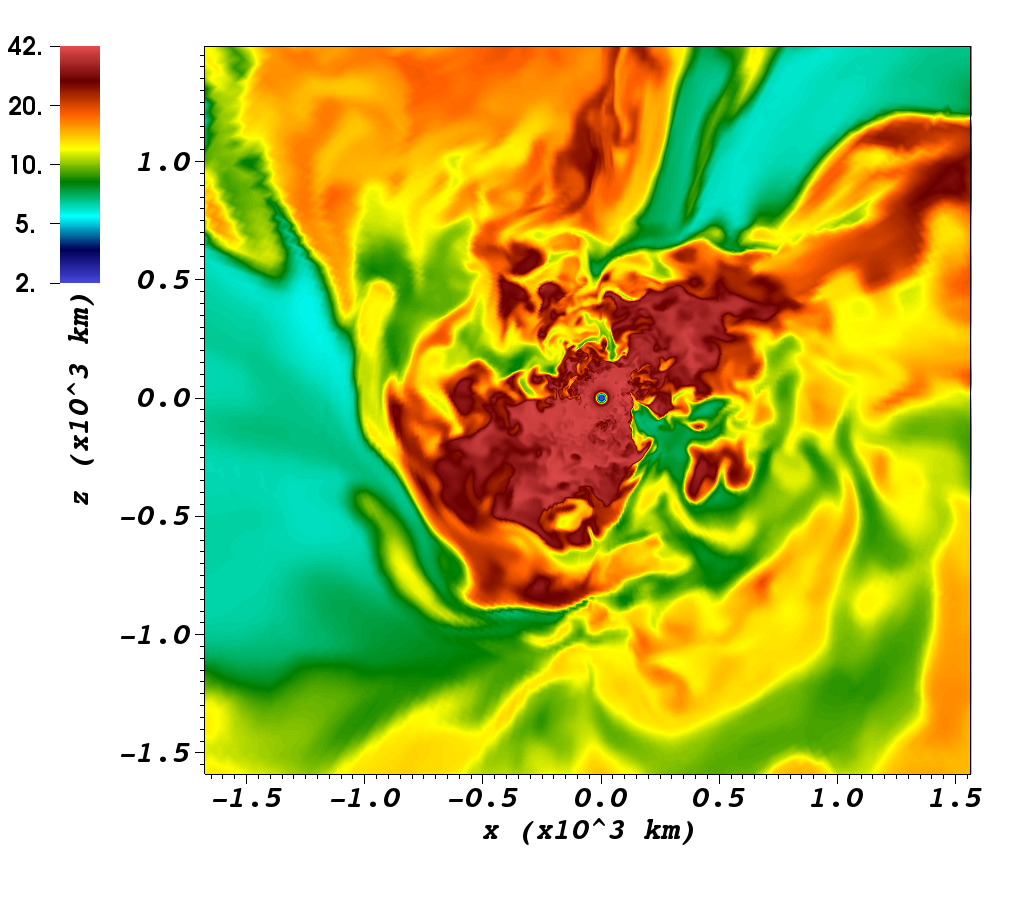}
\includegraphics[width=0.48 \textwidth,trim={0.5cm 1cm 1cm 1cm},clip]{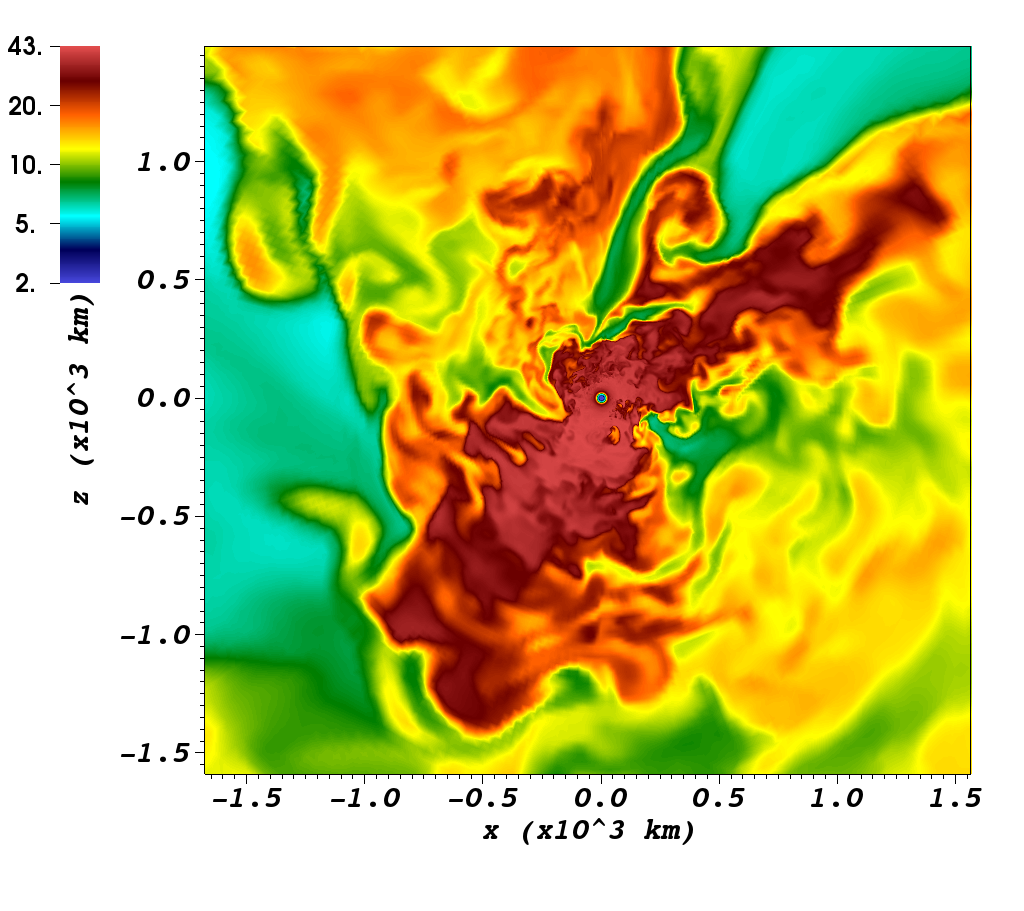}\\
\caption{Entropy $s$ in units of $k_\mathrm{b}/\mathrm{nucleon}$
on 2D slices for model s11.8 at post-bounce times of
$798\, \mathrm{ms}$ (top left),
$911\, \mathrm{ms}$ (top right),
$1045\, \mathrm{ms}$ (bottom left),
and $1094\, \mathrm{ms}$ (bottom right). As time progresses, the
two strong high-entropy outflows in the directions of 11 o'clock and 3'o clock to 6 o'clock are replaced with a new outflow in the 7 o'clock direction.
\label{fig:slices_s118}}
\end{figure*}

\begin{figure*}
    \centering
    \includegraphics[width=\linewidth]{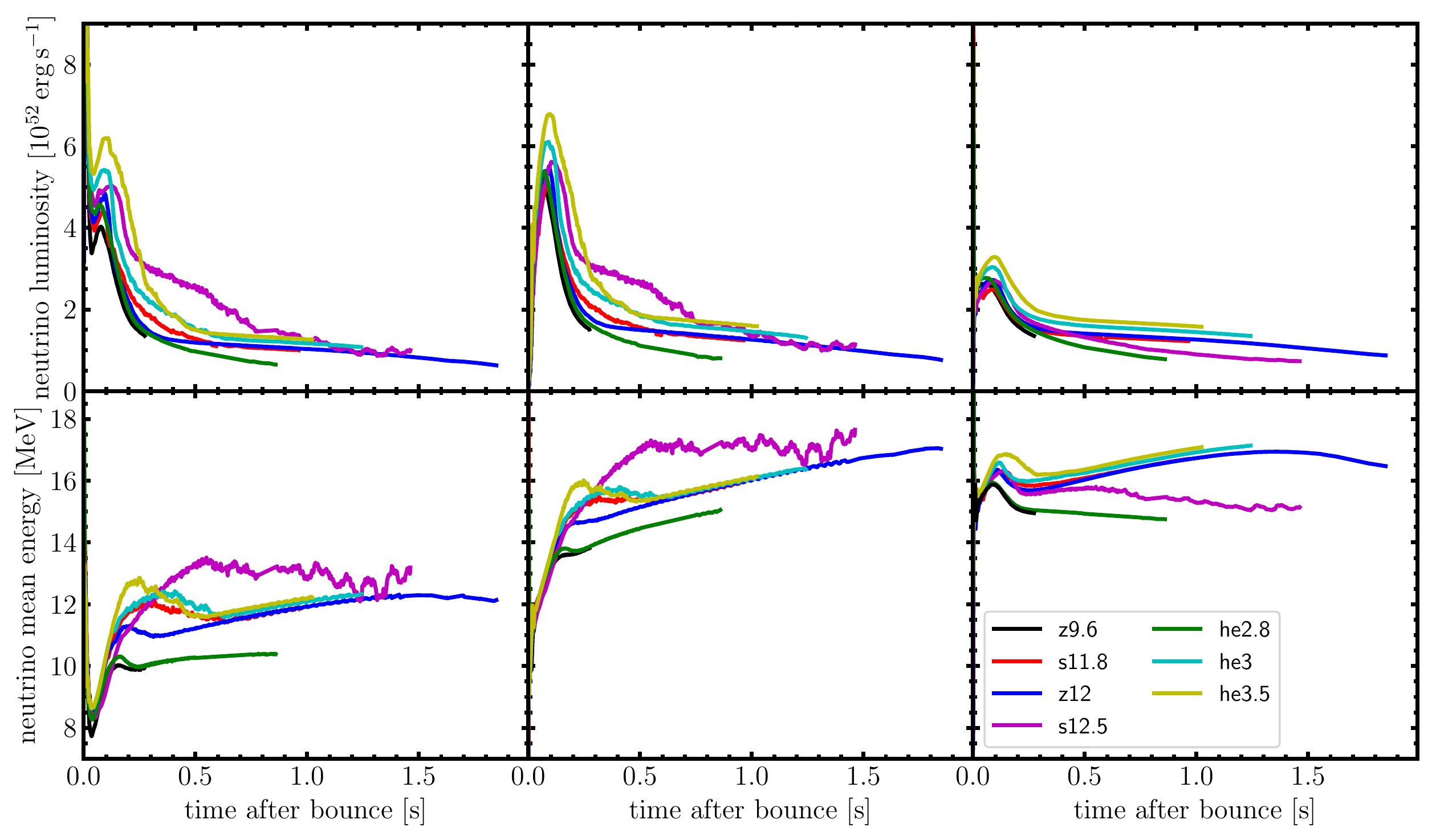}
    \caption{Luminosities (top row) and mean energies (bottom row) for
    electron neutrinos (left column), electron antineutrinos (middle), and heavy flavour neutrinos (right) for all models.
    \label{fig:neutrino}}
\end{figure*}

Shock propagation is fastest for models z9.6 and he2.8, which have very
thin O and C/O shells and hence exhibit the most rapid drop of the
accretion rate.  Residual accretion is thus quickly overwhelmed by the
developing neutrino-heated outflows. Without the supply of fresh
matter at the gain radius, the mass outflow rate
$\dot{M}_\mathrm{out}$ also declines strongly after shock revival, and
the explosion energy therefore essentially plateaus at a low value as
can be seen for model he2.8. The diagnostic energy at the end of these
two simulations is only $1.32 \times 10^{50}\,\mathrm{erg}$ for model z9.6 and $1.12 \times 10^{50}\,\mathrm{erg}$ for model he2.8.

Among the other models, z12 exhibits the lowest accretion rates, but
the accretion phase is much more drawn-out than for z9.6 and
he2.8. This also allows the model to maintain a higher outflow rate
of neutrino-heated matter, and the explosion energy $E_\mathrm{expl}$,
which initially grows at a similar rate as in z9.6 and he2.8,
plateaus later.  By the end of the simulation at $1.847\, \mathrm{s}$,
$E_\mathrm{expl}$ has already reached a value $4.1\times10^{50}\,\mathrm{erg}$;
although the explosion energy has not finally saturated yet, its rate
of increase has already slowed considerably.

Model he3.5 also shows first signs of the explosion energy converging
towards its asymptotic value.  While shock revival occurs somewhat
later at a post-bounce time
of $200 \, \mathrm{ms}$ due to a later infall of the O shell, the
accretion rate also drops quickly thereafter, approaching similarly
low values as for z12 after $\mathord{\sim}700 \, \mathrm{ms}$. At the
end of the simulation, the growth of the explosion energy has slowed
down considerably, and we obtain a final value of $E_\mathrm{expl}=3.66 \times 10^{50}\, \mathrm{erg}$.

Models s11.8 and he3.0 evolve in a remarkably similar way until about
$\mathord{\sim}700 \, \mathrm{ms}$ in terms of their mass accretion
rate, mass outflow rate, shock propagation, and explosion energy. At
that point the models part company with the growth of the explosion
energy in model s11.8 slowing down. It is not clear whether this
already indicates that the explosion energy in s11.8 is nearing
saturation. Although this model shows a stronger decline of
$\dot{M}_\mathrm{acc}$ than he3.5 at late times, the mass outflow rate
$\dot{M}_\mathrm{out}$ is still similar, suggesting that the slower
growth of the explosion energy is not due to a lack of supply of fresh
matter at the gain region.  The slower growth rate is instead due to a
lower average total enthalpy $\bar{h}_\mathrm{tot}$ in the outflows
(Figure~\ref{fig:enthalpy}), which is the main determining factor for
$E_\mathrm{expl}$ along with the mass outflow rate \citep{mueller_15b}. In
model s11.8, $\bar{h}_\mathrm{tot}$ drops significantly below the
typical values of $6\texttt{-}9 \, \mathrm{MeV}/\mathrm{baryon}$ for
the other models with sustained accretion. In model he2.8, a similar
drop marks the transition from the initial explosion phase to the
incipient neutrino-driven wind phase, but a close inspection of the
multi-dimensional flow dynamics in model s11.8 points towards a
different reason for the drop in
$\bar{h}_\mathrm{tot}$. Figure~\ref{fig:slices_s118} shows that this drop
coincides with a significant realignment of the downflow and
outflow geometry. Initially, the model is characterised by a strong
outflow in the 3 o'clock to 6 o'clock direction (top left panel in Figure~\ref{fig:slices_s118}, $798\, \mathrm{ms}$ after bounce). During the
next few hundreds of milliseconds, a downflow from the 9 o'clock
direction intrudes into and mixes with this outflow (top right panel
in Figure~\ref{fig:slices_s118}, $911\, \mathrm{ms}$).  Hence much of the
ejected neutrino-heated material is diluted with cold matter from the
downflows, which lowers the average energy and enthalpy of the
outflow.  Later on (bottom row in Figure~\ref{fig:slices_s118}, $1045\,
\mathrm{ms}$ and $1094\, \mathrm{ms}$), a new outflow of high-entropy
material develops into the 7 o'clock direction.  Due to the limited
simulation time, we cannot exclude that this new outflow grows further
and reinvigorates the growth of the explosion energy. The
reorientation of the outflow bears some resemblance to the phenomenon
of outflow quenching in 2D simulations \citep{mueller_15b}, albeit in
less dramatic form. It suggests that the energetics of 3D models after
shock revival can still exhibit some degree of stochasticity and is
not determined by bulk parameters like the total accretion rate, and
the neutron star mass and radius alone.

Model s12.5 is characterised by significantly higher accretion rates
than the other cases, and the drop in $\dot{M}_\mathrm{acc}$
associated with the infall of the O shell is not very
pronounced. Despite rather strong convective seed perturbations in the
O shell, this delays shock revival to about $500 \, \mathrm{ms}$ after
bounce.  During the explosion phase, the mass outflow rate
$\dot{M}_\mathrm{out}$ remains low compared to the mass accretion
rate; the model is the only one that still exhibits a positive net
accretion rate onto the PNS. This results in a slow
growth of the explosion energy. Without additional simulations, it is
not possible to definitively pin down the reasons behind the rather
tepid explosion of model s12.5 as compared to he3.0, he3.5, and z12,
but it is likely that a combination of factors contribute. A
comparison of the neutrino emission in the different simulations
(Figure~\ref{fig:neutrino}) reveals that s12.5 only exhibits modestly
higher electron flavour luminosities and mean energies than the other
progenitors in spite of considerably higher accretion rates. This is
especially true beyond the first second, when the electron neutrino
and antineutrino luminosity even drops below he3.0, he3.5, and z12. In
the models that include strangeness corrections and nucleon
correlations, somewhat faster diffusion of neutrinos from the outer
layer of the PNS helps to maintain higher
neutrinospheric temperatures at these rather late times.  During the
first $\mathord{\sim}0.5 \,\mathrm{s}$ of the explosion, the neutrino
luminosities and mean energies are, however, noticeably higher than in
the other models, and should in principle allow for significantly
stronger heating and a higher mass outflow rate, which is not
observed. The explosion geometry may partly explain why neutrino
heating is less efficient at driving outflows in this simulation. All
of the other models are characterised some degree of bipolarity at the
early stages of the explosion, either with two similarly prominent
outflows (model z12) or with a strong and a subdominant outflow in the
opposite direction (z9.6, s11.8, he2.8, he3.0, he3.5).  In model
s12.5, by contrast, the explosion is clearly unipolar from early
times. Accretion proceeds mostly through a very broad downflow that
covers almost one entire hemisphere and undergoes very little turbulent
braking and turbulent mixing before reaching the PNS. 
The narrower downflows in the other models dissolve more readily
further away from the PNS, which implies a lower burden
for the re-ejection of the accreted matter and thus allows higher
outflow rates \citep{mueller_15b,mueller_17}.

\begin{figure}
    \centering
    \includegraphics[width=\linewidth]{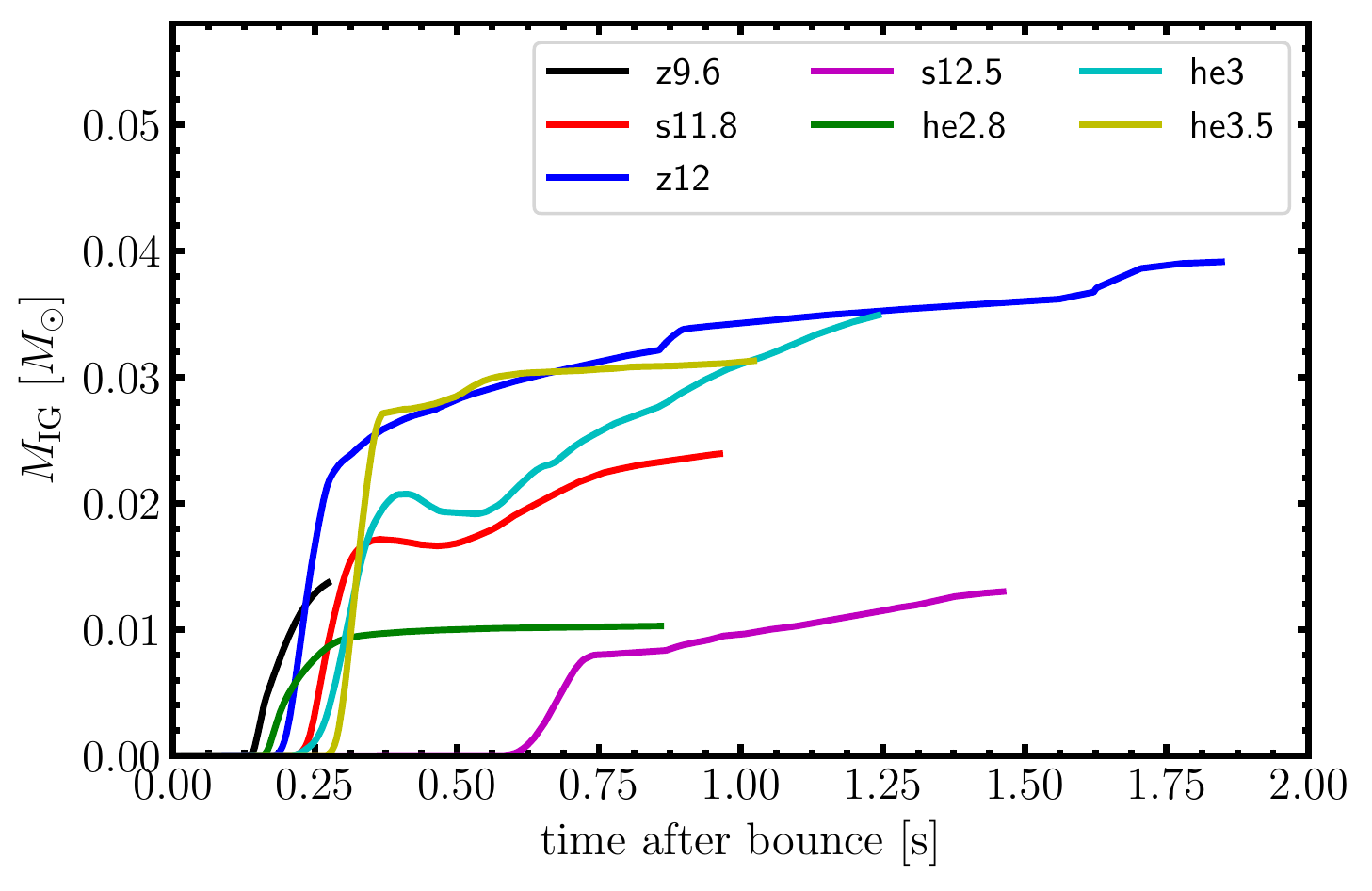}
    \caption{Total mass of iron-group elements in the ejecta (defined
    as the material that is nominally unbound at any given time) as a
    function of post-bounce time.
    \label{fig:ejecta}}
\end{figure}

\subsection{Ejecta Composition}
\label{sec:ejecta}
Due to our simple flashing treatment and the approximations in the
neutrino treatment, we can only draw limited conclusions on the inner
ejecta of the simulated explosions. In particular, uncertainties in
the electron fraction $Y_\mathrm{e}$ translate into an uncertainty in
the composition of the iron group ejecta made by (partial)
recombination of the neutrino-processed ejecta as discussed previously
in \citet{mueller_17}.  Models with more sophisticated neutrino
transport tend to predict predominantly proton-rich outflows with
$Y_\mathrm{e}>0.5$ \citep{pruet_06,froehlich_06,mueller_12a,wanajo_18,vartanyan_18}, although significant
amounts of neutron-rich material can be produced in rapidly developing
explosions. Under ``normal'', proton-rich conditions, $^{56}\mathrm{Ni}$ is the
predominant nucleus in the iron group ejecta \citep{hartmann_85}.

Following our approach in \citet{mueller_17,mueller_18}, we therefore
consider the total mass $M_\mathrm{IG}$ of unbound iron group material
as an estimator for the mass of radioactive nickel
(Figure~\ref{fig:ejecta}). In line with previous 3D simulations
\citep{melson_15a,mueller_17}, the models are characterised by a
steep rise of $M_\mathrm{IG}$ immediately at the onset of the
explosion, which stems from the recombination of shocked,
photodisintegrated material that never makes it close to the
PNS and/or from iron group material from explosive
burning that is entrained by the neutrino-heated bubbles.  Later on,
the neutrino-driven outflows contribute further iron group material
made by partial recombination at a smaller rate.

Like the explosion energies, the iron group masses have not fully
converged yet, although the increase of $M_\mathrm{IG}$ has already
flattened considerably in he2.8, he3.5, s11.8, z12, and s12.5. The
preliminary values at the end of the simulations lie between $0.01
M_\odot$ and $0.04 M_\odot$, and there is a clear correlation between
explosion energy and nickel mass. This is broadly compatible with
observationally inferred values for low-mass supernovae 
\citep{chugai_00,fraser_11,pejcha_15c,lisakov_18}
and the well-established observational correlation between
$E_\mathrm{expl}$ and the nickel mass \citep{hamuy_03,pejcha_15c}.

\begin{figure}
    \centering
    \includegraphics[width=\linewidth]{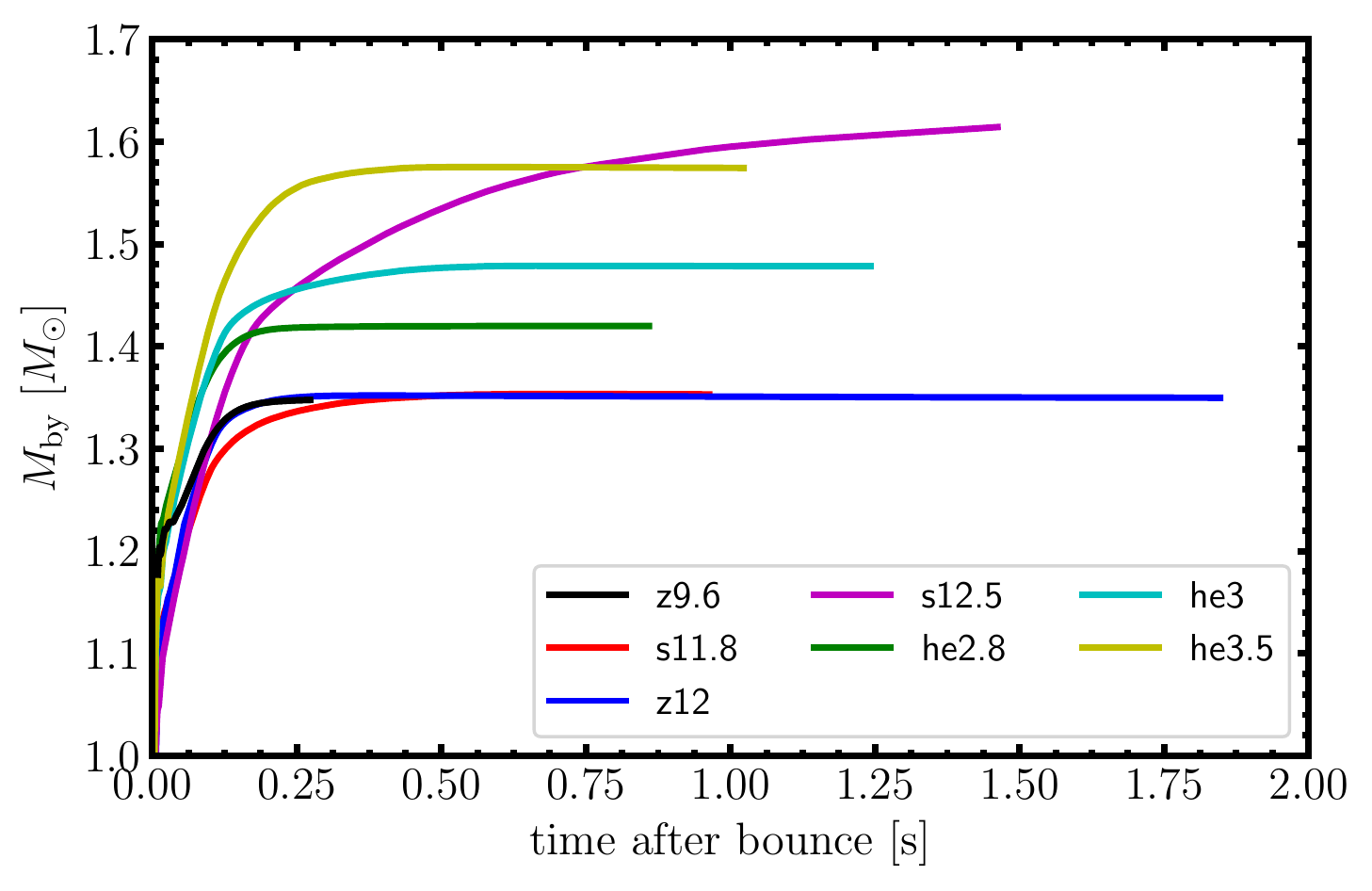}
    \caption{Baryonic PNS masses $M_\mathrm{by}$ as a function
    of post-bounce time. Except for model s12.5, $M_\mathrm{by}$ has
    essentially reached its final value.
    \label{fig:masses}}
\end{figure}

\begin{figure}
    \centering
    \includegraphics[width=\linewidth]{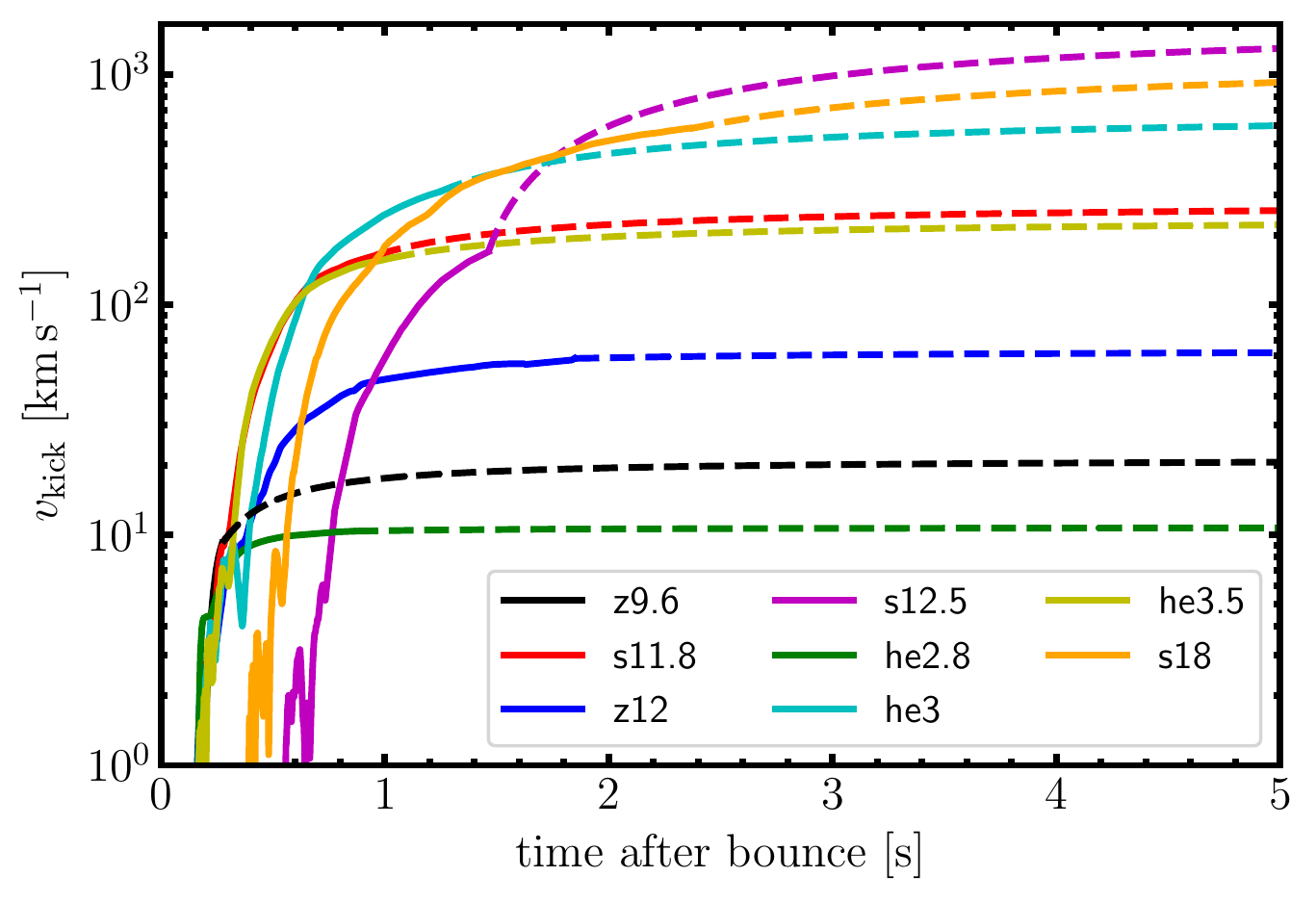}
    \caption{PNS kick velocity $v_\mathrm{kick}$ as a function of post-bounce time for all models. During the simulated evolution the kick is computed from the momentum of the ejecta
    invoking momentum conservation (solid curves). In addition to our
    seven low-mass progenitors, the $18 M_\odot$ model (s18) of
    \citet{mueller_17} is also included.
    In order to
    extrapolate the kick beyond the final simulation time (dashed curves),
    we use Equation~(\ref{eq:extrapolation}), which assumes that
    the late evolution of the kick is dominated by the gravitational
    tug of the ejecta, and that the ejecta expand roughly homologously.
    Except for model s12.5, this yields a very smooth extrapolation.
    \label{fig:kicks}}
\end{figure}

\begin{figure}
    \centering
    \includegraphics[width=\linewidth]{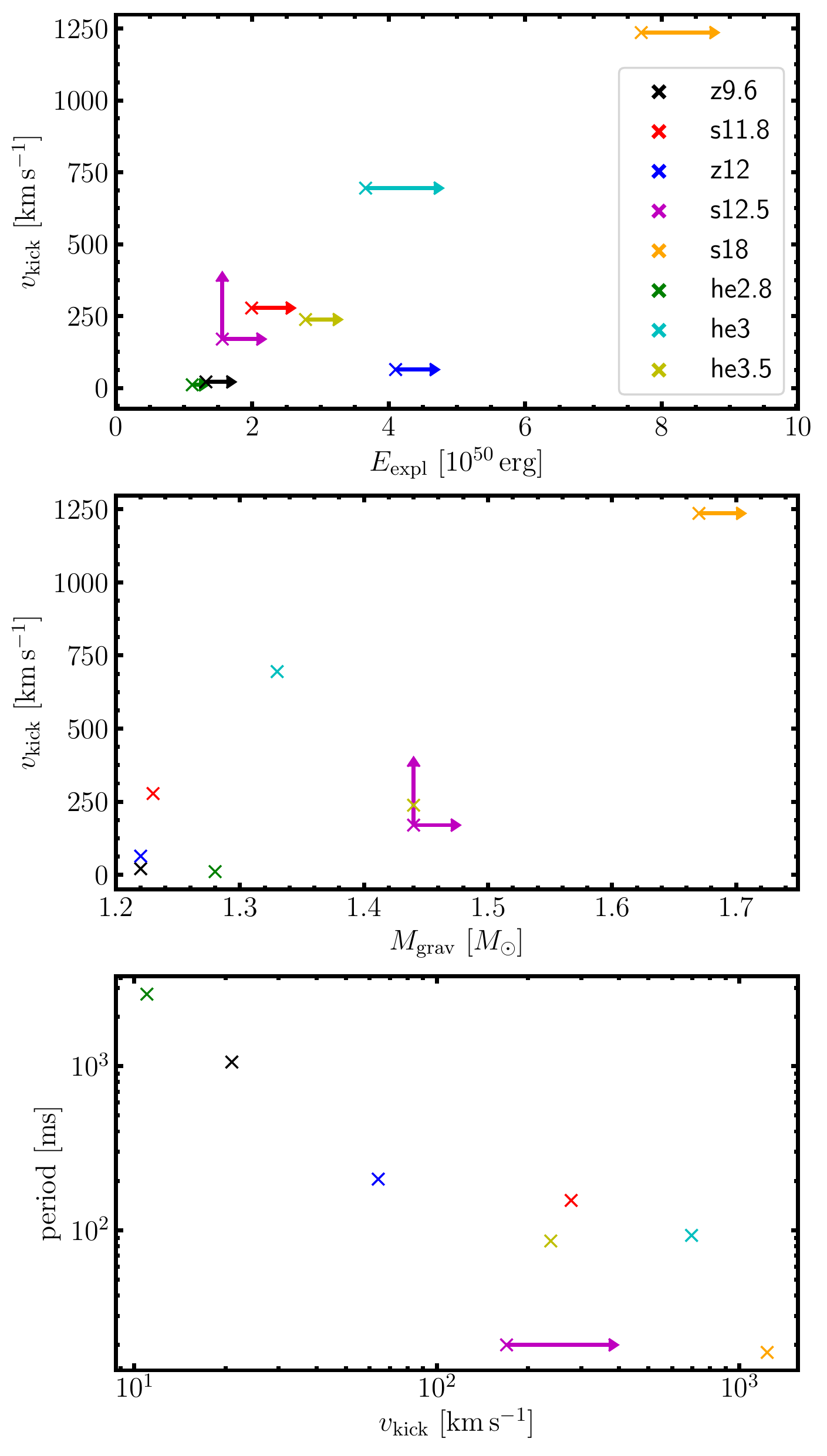}
    \caption{Correlations between
    explosion energy
    $E_\mathrm{expl}$ and the (extrapolated) neutron star
    kick $v_\mathrm{kick}$ (top panel), between
    the gravitational neutron star mass $M_\mathrm{grav}$
    and $v_\mathrm{kick}$ (middle panel),
    and between $v_\mathrm{kick}$ and the spin period
    (bottom panel). The plots include the 
    $18 M_\odot$ model (s18) of \citet{mueller_17} in addition to
    our seven low mass models. Since the explosion
    energy is still evolving in all of the models, the plotted
    values are only lower limits, the same is true
    for $M_\mathrm{grav}$ in models s12.5 and s18 and $v_\mathrm{kick}$
    in model s12.5. The length of the
    arrows has no firm quantitative meaning, but gives a subjective estimate for the further growth of these uncertain quantities.
    \label{fig:correlations}}
\end{figure}

\begin{figure}
    \centering
    \includegraphics[width=\linewidth]{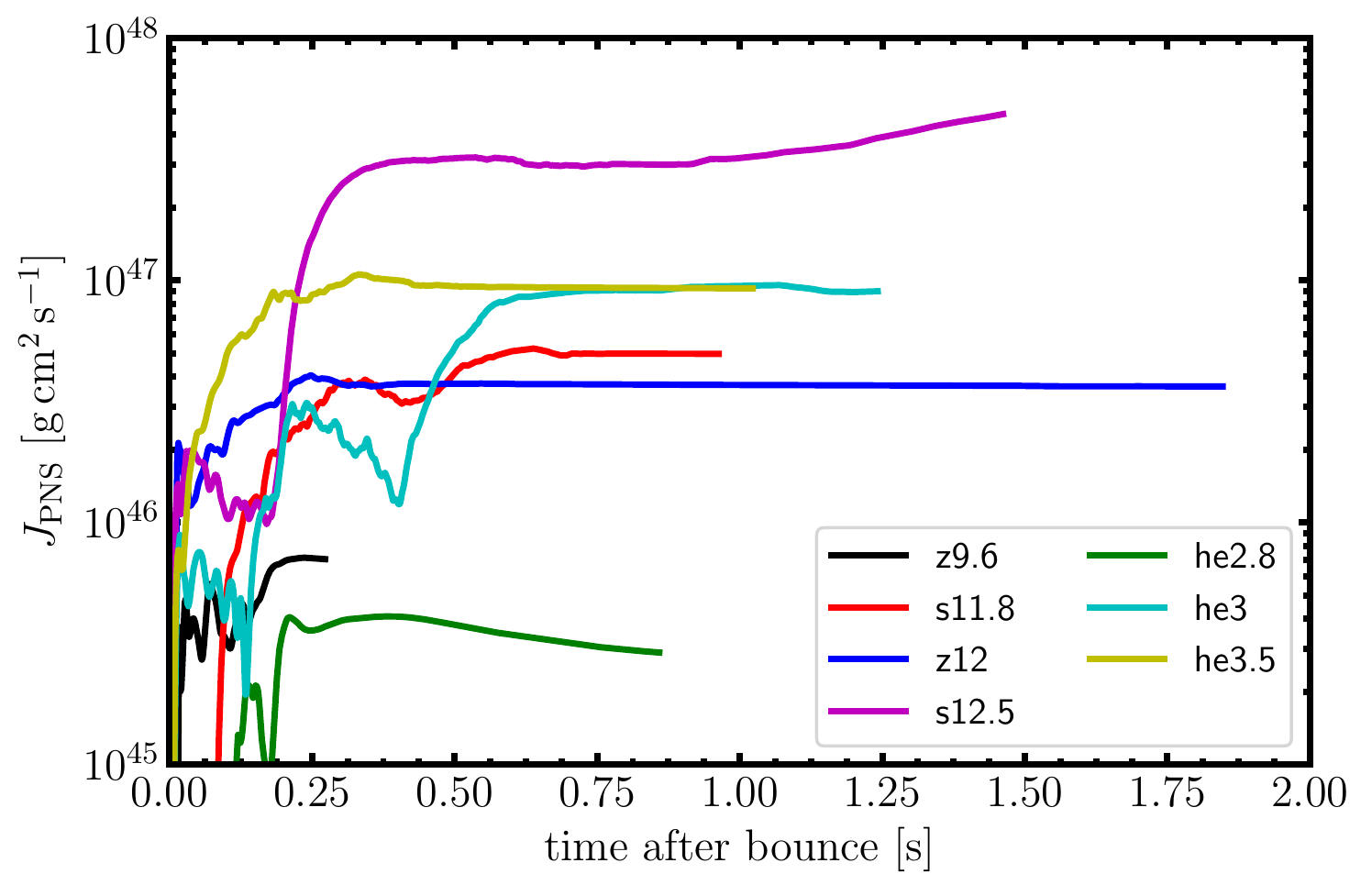}
    \caption{Angular momentum $J_\mathrm{PNS}$ advected onto
    the PNS computed according to Equation~(\ref{eq:jpns}).
    \label{fig:j_pns}}
\end{figure}

\begin{figure}
    \centering
    \includegraphics[width=\linewidth]{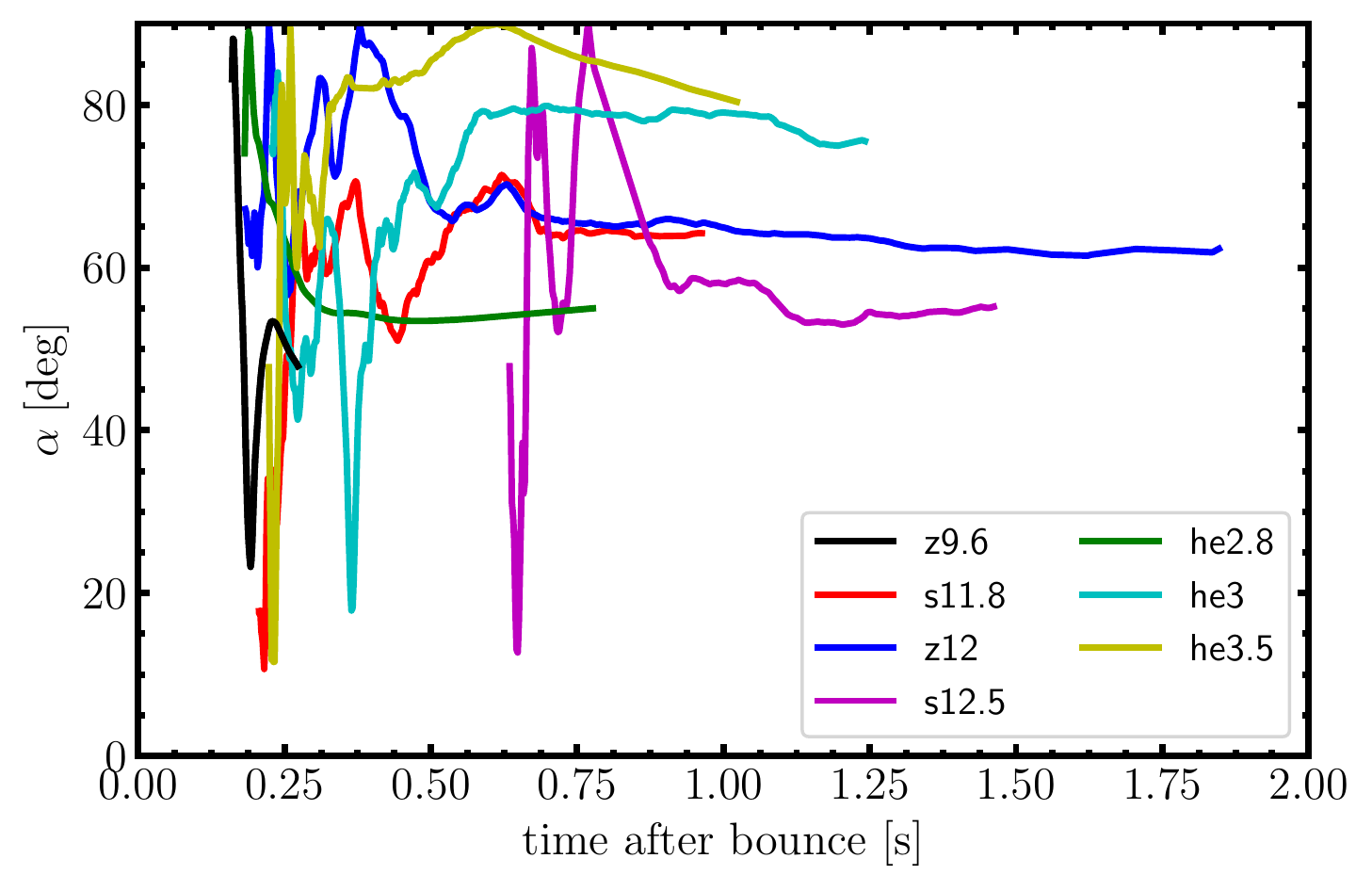}
    \caption{Evolution of the angle $\alpha$ between the spin and kick direction. Since this angle fluctuates rapidly as long as the spin and kick are small, we only evaluate $\alpha$ once the diagnostic energy has reached a significant positive vallue.
    \label{fig:alignment}}
\end{figure}

\subsection{Neutron Star Properties}
Except for the case of s12.5, the uncertainties in the final neutron
star properties are considerably smaller than for the explosion
energies and nickel masses.

\subsubsection{Neutron Star Masses}
Figure~\ref{fig:masses} shows the baryonic mass $M_\mathrm{by}$ of the
PNS for all seven simulations. In six of the models
(z9.6, s11.8, z12, he2.8, he3, he3.5), $M_\mathrm{by}$ has essentially
asymptoted to its final value, and actually decreases at a very small
rate since the outflow rate already exceeds the accretion
rate. Barring the possibility of late-time fallback, the
values at the end of the runs can therefore be taken as upper limits for
these six models.

Approximately correcting for the binding energy of cold neutron stars following \citet{lattimer_01} to obtain gravitational masses $M_\mathrm{grav}$,
\begin{equation}
  M_\mathrm{grav} \approx M_\mathrm{by} - 0.084 M_\odot
  \left(\frac{M_\mathrm{grav}}{M_\odot}\right)^2,
\end{equation}
we find values between $1.22 M_\odot$ for z9.6 and 
$1.44 M_\odot$ for s12.5 (which is still a lower limit
for this particular model). 

Even though a comparison with observed neutron star masses is
complicated by intricacies of binary evolution like mass transfer and
binary breakup, it is noteworthy that these gravitational masses fall
nicely within the range of measured masses ($\sim 1.17\texttt{-}1.6 M_{\odot}$) in double neutron star
systems \citep{martinez_15,oezel_16,tauris_17,ferdman_18}.  
This suggests that even the rather high neutron star mass of $\mathord{\sim} 1.67\,M_\odot$ for the $18\,M_\odot$
explosion model of \citet{mueller_17} is merely an outlier among the double neutron star systems and may
otherwise be explained by the rarer observed cases of high neutron star birth masses ($\mathord{>}1.7 M_{\odot}$) in binary systems in general \citep{tauris_11}. Hence, we conclude that our current 3D neutrino-driven explosion
models are in fine agreement with observed neutron star masses.

\subsubsection{Neutron Star Kicks}
Estimating the final values of the neutron star kick is somewhat more
complicated. Following previous studies, we calculate the kick
velocity $v_\mathrm{kick}$ by evaluating the total momentum of the
ejecta and invoking momentum conservation \citep{scheck_06},
\begin{equation}
  \label{eq:vkick}
  \mathbf{v}_\mathrm{kick}=-
  \frac{1}{M_\mathrm{grav}}
  \int_\mathrm{ejecta} \mathbf{S} \,\ud \tilde{V}.
\end{equation}
Here $\mathbf{S}$ and $\ud\tilde{V}$ are the momentum density and volume
element including relativistic correction terms (which are immaterial
in practice, however). The contribution of anisotropic neutrino
emission to the kick is by far subdominant
\citep{mueller_17,gessner_18,mueller_18}, and is therefore neglected.

Figure~\ref{fig:kicks} shows the the evolution of $v_\mathrm{kick}$
for all seven models. Although the acceleration of the PNS has already slowed down in most of the models, it has clearly not
yet reached its final asymptotic value.  Fortunately, however, the ongoing
acceleration of the PNS is mostly due to the asymmetric
gravitational ``tug'' \citep{wongwathanarat_13} of the early ejecta.  Except for
model s12.5, asymmetries in the downflows and outflows closer to the
PNS no longer strongly affect the evolution of the kick
for two reasons: Not only has the accretion rate already dropped
considerably so that there is little mass around close to the
PNS to exert a gravitational tug; the orientation and
strength of downflows and outflows is also quite variable at small
radii so that there is little net acceleration of the PNS over time.

This allows us to tentatively extrapolate the kick to its asymptotic
value for most of the models. With long-range gravitational forces
dominating the evolution of the kick at late times, the acceleration
$\mathbf{a}_\mathrm{kick}=\dot{\mathbf{v}}_\mathrm{kick}$ of the
PNS is essentially given by
\begin{equation}
\mathbf{a}_\mathrm{kick}
=
\int_\mathrm{ejecta}
    \frac{G \rho M_\mathrm{grav} \mathbf{r}}{r^3}\,\ud V.
\end{equation}
Since the geometry of the early ejecta is already quite stable at late
times, and since they only undergo modest deceleration over time
scales of seconds, we can approximate their expansion as roughly
self-similar, which implies that the acceleration scales inversely to
the square of the scale factor $\mathcal{S}$ at different times.  The
scale factor $\mathcal{S}$ is roughly proportional to the elapsed time
since the onset of the explosion. Since the definition of the
explosion time is somewhat ambiguous, and since most of our models
explode early and are evolved to rather late times anyway, we
approximate the time dependence of the acceleration as
\begin{equation}
\label{eq:extrapolation}
\mathbf{a}_\mathrm{kick}(t)
\approx \left(\frac{\mathcal{S}(t)}{\mathcal{S}(t_0)}\right)^{-2} \mathbf{a}_\mathrm{kick}(t_\mathrm{fin})
\approx
\left(
\frac{t-t_\mathrm{bounce}}{t_\mathrm{fin}-t_\mathrm{bounce}}\right)^{-2}
\mathbf{a}(t_\mathrm{fin}),
\end{equation}
where $t_\mathrm{bounce}$ and $t_\mathrm{fin}$ are the time of bounce
and the final simulation time. We obtain $\mathbf{a}(t_\mathrm{fin})$
by averaging the gravitational force exerted onto the PNS  over an interval of $50\texttt{-}200\,\mathrm{ms}$ before the end
of each simulation.  Equation~(\ref{eq:extrapolation}) can then be
integrated analytically.
In most models, this approximation already gives
a good fit during the late phases of the simulations,
which provides addition confidence in its validity.
Compared to the extrapolation procedure of
\citet{scheck_06}, who merely applied a constant acceleration
$\mathbf{a}(t_\mathrm{fin})$ over a manually specified time frame,
Equation~(\ref{eq:extrapolation}) furnishes a less ambiguous
extrapolation method.

Extrapolated values for $v_\mathrm{kick}$ are shown as dashed curves
in Figure~\ref{eq:vkick}, and the asymptotic values are listed in
Table~\ref{tab:expl_prop}.  With the exception of model s12.5,
Equation~(\ref{eq:extrapolation}) indeed provides a smooth
extrapolation of the kick velocities. The inferred final values range
from $11\texttt{-}21 \, \mathrm{km}\, \mathrm{s}^{-1}$ for he2.8 and
z9.6 to $695\, \mathrm{km}, \mathrm{s}^{-1}$ for he3. The values at
the low-mass end of the spectrum of single-star and (ultra)stripped-envelope
progenitors of iron-core collapse supernovae are thus of the same order as for ECSNe \citep{gessner_18}. 
This is due to the structural similarity of these models
to ECSN progenitors discussed in Section~\ref{sec:progenitors}:
Models with a very small helium core mass exhibit
a steep density gradient outside the Si core (Figure~\ref{fig:profiles}),
which implies that only a small amount of mass can become involved
in aspherical fluid motions after shock revival, and hence the
gravitational tug on the PNS remains weak.\footnote{\citet{janka_17}
and \citet{gessner_18} have also framed the discussion of
the explosion dynamics of ECSNe in terms of the progenitor compactness,
but compactness is not the ideal basis for understanding
the peculiar explosion dynamics and the low kicks in this regime.
What is relevant to the explosion dynamics and the kicks is rather the mass
within a few hundreds (for ECSN-like models) to thousands of kilometres
of the edge of the Si core, whereas the compactness is nothing
but a measure of the radius of one specified mass shell.}

The kicks of the more massive
progenitors (producing more massive metal cores) appear broadly compatible with the observed distribution of
neutron star kicks \citep{hobbs_05,faucher_06,ng_07}. Again, our simulations indicate that
the rather high value found for the $18 M_\odot$ model of
\citet{mueller_17}  --- $600\, \mathrm{km} \, \mathrm{s}^{-1}$ at the end
of the simulation or $1236\, \mathrm{km} \, \mathrm{s}^{-1}$ after
using our extrapolation procedure ---  is an outlier rather than a symptom
of a generic problem of self-consistent neutrino-driven explosion
models. It is remarkable that the extrapolated value for the
model of \citet{mueller_17} would place it just around the highest
observed kick velocities \citep{cordes_93,chatterjee_05,tomsick_12,tauris_15b}.

It is interesting to consider models he2.8 and he3.5 on their own.
Since these two models are representative examples for low- and relatively high-mass
metal cores ($1.47 M_\odot$ and $1.81 M_\odot$, respectively) that can form in  
ultra-stripped progenitors\footnote{Note, strictly speaking, ultra-stripped progenitors
are defined as exploding stars with an envelope mass of $\mathord{\la} 0.2M_\odot$ \citep{tauris_15}.}, 
we can assume that they allows us to probe the range of kicks
that can be achieved in this evolutionary channel. It is noteworthy that model
he3.5 develops a considerably larger kick ($159 \, \mathrm{km}\,
\mathrm{s}^{-1}$) than has been found so far in simulations of
ultra-stripped supernovae \citep{suwa_15,mueller_18}. Thus, we
estimate that kicks for ultra-stripped explosions should fall roughly
between $\sim 10\,\rm{km\,s}^{-1}$ and (at least) $\sim 200\,\rm{km\,s}^{-1}$, 
compatible with the properties of double neutron star systems that 
formed via the ultra-stripped channel.
Indeed, \citet{tauris_17} found evidence for a range of ultra-stripped supernova kicks 
(producing the second-formed neutron star in known double neutron star systems) 
with a majority of small kicks $<50\,{\rm km\,s}^{-1}$ \citep[see also][]{beniamini_16} and a few systems with
large kicks \citep[see also][]{fryer_97,wex_00}. For example, PSRs B1534+12 and B1913+16 were found to have experienced neutron star
kicks of $175\texttt{-}300\,{\rm km\,s}^{-1}$ and $185\texttt{-}465\,{\rm km\,s}^{-1}$, respectively.
Based on our simulations presented here, it is possible that explosions of 
ultra-stripped stars with metal cores $\mathord{\ga} 2.0\,M_\odot$ may result
in neutron star kicks in excess of $200\,{\rm km\,s}^{-1}$. This will be investigated 
in future simulations.

It has recently been proposed that there could be a strong correlation
between neutron star kicks and progenitor mass
\citep{bray_16,bray_18}, or rather with the explosion energy and
the mass $M_\mathrm{ej}$ of the shocked ejecta around the time when
the kick is determined \citep{janka_17,vigna_gomez_18}.
In terms of
$E_\mathrm{expl}$, $M_\mathrm{ej}$, $M_\mathrm{grav}$, and an
anisotropy parameter $\alpha_\mathrm{ej}$, one expects
\citep{vigna_gomez_18}
\begin{equation}
\label{eq:kick_scaling}
    v_\mathrm{kick}
    =\alpha_\mathrm{ej} \frac{\sqrt{E_\mathrm{expl} M_\mathrm{ej}}}{M_\mathrm{grav}}.
\end{equation}
Since it can be argued that $E_\mathrm{expl}\propto M_\mathrm{ej}$
\citep{janka_17}, and since supernova explosion energies vary
considerably more than neutron star masses, this would imply
$v_\mathrm{kick} \propto \alpha_\mathrm{ej} E_\mathrm{expl}$. 
If, as argued by \citet{vigna_gomez_18}, the anisotropy parameter is more or less universal over a wide range of progenitor masses, one would
expect the kick velocity to be strongly correlated with the explosion
energy.
Similarly, \citet{tauris_17} argue that one expects the kick velocity to be correlated with the resulting neutron star mass (see below).
Simulations have not been able to properly address these
hypotheses so far, however. While systematic studies of the
gravitational tug-boat mechanism have to some degree investigated the
dependence of the kick on explosion energy using parameterised
explosion models \citep{scheck_06,wongwathanarat_13,gessner_18}, this
approach is not problematic for determining correlations
between the kick and other explosion properties: Unless carefully
calibrated, it will reveal correlations that are due to variations in
parameters of the neutrino-driven engine (which would not vary in
nature) rather than correlations due to variations in progenitors
structure.  Our self-consistent models allow us to better address
possible correlations between the kick and the explosion energy
(though caveats remains because of the small sample size and small
differences in the neutrino interaction rates between the
simulations).

In the top panel of Figure~\ref{fig:correlations}, we plot the extrapolated final kick
velocity (where applicable) against the diagnostic explosion energy at
the end of our simulation for the low-mass models presented in this
paper, and also add the $18 M_\odot$ model of \citet{mueller_17}. The
values for the explosion energy are to be taken only as lower limits,
but for most of the models (except he3), one would not expect a
further increase by more than a few tens of percent, so that the
preliminary values of $E_\mathrm{expl}$ are a good indicator of the
ordering of the final explosion energies.

Our simulations appear compatible with a moderately strong correlation
between kick velocity and explosion energy or, alternatively, 
the mass of iron-group ejecta, which is tightly linked to the explosion
energy as discussed in Section~\ref{sec:ejecta} and can be seen
as a proxy for the relevant ejecta mass $M_\mathrm{ej}$
in Equation~(\ref{eq:kick_scaling}). Model z12 is a significant
exception, however. It is the most energetic model among the low-mass
cases considered in this study, yet it yields a low kick velocity of
only $64\, \mathrm{kms}\, \mathrm{s}^{-1}$. This is naturally
explained by the explosion geometry of this model, which stands out
from the other cases as clearly bipolar rather than unipolar
(Figure~\ref{fig:slices}), so that the net gravitational tug remains
close to zero.

Whether or not there is a tight correlation between $v_\mathrm{kick}$
and $E_\mathrm{expl}$ depends on whether such bipolar explosions are
frequent or not, and our current models do not permit any conclusions
on this question. It is possible that the bipolar explosion geometry
of model z12 is an artefact of the older mesh coarsening scheme of
\citet{mueller_15b}, which may favour explosions aligned with the
grids axis in the absence of strong large-scale perturbations in the
progenitor. However, model s11.8, which
has also been simulated using the
old mesh coarsening scheme, casts doubt on this explanation. Its
explosion is clearly not aligned with the grid axis even though the
progenitor does not exhibit violent large-scale convection in the O
shell either that could imprint a preferred geometry onto the model
and swamp grid artefacts. It is therefore equally plausible that 
bipolar explosions as in model z12 are physical and reasonably
frequent. Moreover, the current models do not include rotation, and
even slow rotation could lead to a more bipolar flow geometry. The
small body of self-consistent 3D explosion models in the literature
does not provide much further guidance either.  Only few models have
been evolved sufficiently far by other groups to show the emerging
explosion geometry
\citep{takiwaki_14,melson_15a,melson_15b,lentz_15,ott_18,vartanyan_18},
and although there is a preponderance of unipolar explosions, examples
of more bipolar explosions are also found (e.g.\ the $15 M_\odot$
model of \citealt{ott_18}). Moreover, a number of those simulations
studied more massive progenitors where the spiral mode of the SASI
imprints a strong $\ell=1$ mode onto the flow.

At this point, the evidence thus allows us to conclude only that there
is at least a loose correlation between $E_\mathrm{expl}$ and
$v_\mathrm{kick}$. It is plausible that the achievable kick velocity
indeed scales roughly linearly with $E_\mathrm{expl}$, while there is
considerable scatter below this upper limit. Whether the distribution
of kicks below this limit is strongly top-heavy with a few outliers
with low kicks, more uniform, or whether there is even a bimodality
between unipolar and bipolar explosions as speculated by
\citet{scheck_06} will need to be investigated with a considerably
larger sets of supernova models, and detailed studies of neutron stars
within supernova remnants may also shed light on this question
\citep{katsuda_18}.

\citet{tauris_17} found indications of an empirical correlation between
the kick and the neutron star mass in their analysis of double neutron star
systems and also presented qualitative arguments to support such a correlation. 
In our simulations, however, we find only a loose correlation between those two quantities
(middle panel of Figure~\ref{fig:correlations}). It is to be expected that this correlation
is weaker than that between $E_\mathrm{expl}$ and $v_\mathrm{kick}$. 
The correlation between $E_\mathrm{expl}$ and $v_\mathrm{kick}$ directly
reflects the physics of hydrodynamical kicks and comes about because
$E_\mathrm{expl}$ and $v_\mathrm{kick}$ are both intimately linked
to the mass of ejected neutrino-heated material \citep{janka_17,vigna_gomez_18}.
By contrast, the weaker correlation between $M_\mathrm{grav}$ and $v_\mathrm{kick}$
is likely a secondary consequence of a loose correlation between $M_\mathrm{grav}$
and $E_\mathrm{expl}$ \citep{mueller_16a}: Progenitors with a higher silicon core mass
(which mostly determines $M_\mathrm{grav}$) tend to explode more energetically
because they also tend to have denser and more massive oxygen shells and hence
experience stronger, more sustained accretion after shock revival.  Since this 
correlation is not a tight one in the stellar evolution models, and since it is
further compounded by scatter in the relation between $E_\mathrm{expl}$ and  
$v_\mathrm{kick}$, one may only expect a weak general correlation between $M_\mathrm{grav}$ and  
$v_\mathrm{kick}$ from the theoretical point of view.
As a sub-population, however, ultra-stripped supernovae may exhibit a stronger correlation 
between $M_\mathrm{grav}$ and $v_\mathrm{kick}$. Future simulations and further empirical data is 
needed to confirm or reject this hypothesis.

\subsubsection{Neutron Star Spins}
Since we treat the interior of the PNS as spherically
symmetric and non-rotating in the simulations, we follow
\citet{wongwathanarat_10b,wongwathanarat_13} and calculate its angular
momentum $\mathbf{J}_\mathrm{PNS}$ by integrating the flux of angular
momentum through a sphere of radius $r_0=50 \, \mathrm{km}$ around the
origin,
\begin{equation}
\label{eq:jpns}
  \frac{ \ud \mathbf{J}_\mathrm{PNS}}{\ud t}
  =
  \int \alpha \phi^4 r_0^2   \rho W v_r \mathbf{v} \times \mathbf{r} \,\ud \Omega\
.
\end{equation}
Here $\rho$, $\mathbf{v}$, $v_r$, $W$ are the density, three-velocity,
radial component of three-velocity, and Lorentz factor, and
$\alpha$ and $\phi$ are the lapse function and conformal factor in the
xCFC metric. Since we use non-rotating stellar evolution models, our
analysis only accounts for the spin imparted to the PNS
by asymmetric accretion downflows; however, it is worthwhile to
consider this spin-up during the explosion separately from the poorly
constrained pre-collapse spin periods.

The evolution of $J_\mathrm{PNS}$ is shown in
Figure~\ref{fig:j_pns}. Based on our preliminary values for the
gravitational mass $M_\mathrm{grav}$ and assuming a neutron star
radius of $R=12\, \mathrm{km}$, we compute the corresponding final
final neutron star spin period using the fit formula of \citet{lattimer_05}
for the moment of inertia of cold neutron stars,
\begin{equation}
    I=0.237 M_\mathrm{grav}
    R^2 \left[1+4.2\left(\frac{M_\mathrm{grav} \, \mathrm{km}}{M_\odot R}\right)+90\left(\frac{M_\mathrm{grav} \, \mathrm{km}}{M_\odot R}\right)^4\right].
\end{equation}
The estimated spin periods
are listed in Table~\ref{tab:expl_prop}.

With our non-rotating progenitor models, we obtain a wide range of
spin periods from seconds (z9.6, he2.8) down to $20\, \mathrm{ms}$ for
s12.5, which is similar to the $18 M_\odot$ model of
\citet{mueller_17}. It is quite remarkable that the PNS
angular momentum stabilises quite early in all simulations, and well
before the the explosion energy shows any sign of saturating.  Even in
model s12.5, which only explodes at $0.75\, \mathrm{s}$ after bounce
and still accretes quite heavily at the end of the run, the
$J_\mathrm{PNS}$ does not evolve dramatically any more. In fact,
$J_\mathrm{PNS}$ appears to be set already \textit{before} the onset of
the explosion. This early freeze-out of the proton-neutron star
angular momentum was also observed, though not discussed, in
\citet{mueller_17}, but our models suggest that it could be a rather
generic phenomenon. It is likely explained by the dynamics of the
accretion flow in the explosion phase: The rate of change of
$J_\mathrm{PNS}$ naturally scales with the mass accretion rate onto
the PNS \citep{spruit_98,wongwathanarat_13}, which
drops significantly as the explosion develops in all our models. This
is different from the kick, which can change due to the long-range
force of gravity, and also somewhat different from the explosion
energy. Although the continued growth of the explosion is tied to
ongoing \textit{accretion into the gain region}, much of the accreted
matter no longer settles onto the PNS, but is
re-ejected from relative large turnaround radii \citep{mueller_17} so
that it does not impart any angular momentum onto the compact remnant.

Intriguingly, the spin periods that we obtain for our non-rotating
progenitors due to aspherical accretion alone roughly cover the
observed range of birth period of young pulsars
\citep{faucher_06,perna_08,popov_12,noutsos_13,igoshev_13}.
Although our sample is still too small and too selective to
draw conclusions on the shape of the distribution, it fits the
picture suggested by observations with a skewed distribution
and significant clustering of birth periods below $\mathord{\sim}200 \, \mathrm{ms}$ \citep{popov_12,noutsos_13,igoshev_13}. This
is significantly different from the parameterised explosion models
of \citet{wongwathanarat_13}, which do not compute the diffusive
neutrino flux form the PNS core self-consistently.
These show less spin-up by aspherical
accretion (with typical spin periods of
$\mathord{\gtrsim} 0.5 \, \mathrm{s}$), presumably because the neutrino
flux from the core is somewhat on the high side and quenches
the accretion flow faster than in self-consistent models. 
Naturally, our results do not imply that the PNS spin
is set by the physics of the early explosion phase only. In reality,
the spin of the progenitor core may not be negligible, in particular for
for exploding stars in tight binaries where tidal effects are at work.
It is possible that even the presence of modest amounts of angular
momentum in the progenitor core qualitatively alters the mechanism
of spin-up seen in our models by providing a preferred
axis for the convective flow around the PNS and forcing it
into a non-stochastic flow pattern. However,
for the rotation rates predicted by current stellar evolution
models \citep{heger_05}, one obtains high convective Rossby numbers
$\mathrm{Ro}\gtrsim 10$ in the gain region, which suggests that
rotation is not fast enough to affect the geometry of the convective
flow \citep{mueller_16b}. First 3D simulations of moderately rotating
progenitors \citep{summa_18} appear to confirm this expectation.
Nonetheless, the interplay of stochastic spin-up and progenitor
rotation needs to be investigated further in the future. In addition
to the spin of the progenitor that is neglected in our simulations, another caveat 
concerns fallback, which could alter the PNS spin appreciably during the
late phase of the explosion.

Some indications that the origin of neutron star spins cannot 
be understood based on asymmetric accretion during the first seconds
alone comes from an anti-correlation between kick velocities and spin
periods that emerges for our models (bottom panel of Figure~\ref{fig:correlations}).
Although there is considerable scatter, we see a clear trend towards
shorter spin periods for higher kick velocities, which is again
different from \citet{wongwathanarat_13}. Such a correlation is
not unexpected (and has in fact already been anticipated by 
\citealt{spruit_98}) since the magnitude of both the kick and spin depends on the mass that is involved in overturn motions behind the
shock, which varies considerably between progenitors and thus naturally
accounts for a high degree of covariance between $v_\mathrm{kick}$
and $J_\mathrm{PNS}$.

This correlation is not unproblematic, however. Observationally,
there are examples of short spin periods of young neutron stars with low
kicks; the Crab pulsar is the most prominent example with
a kick of $\mathord{\sim} 160 \, \mathrm{km}\, \mathrm{s}^{-1}$
\citep{kaplan_08,hester_08} and a birth spin period
of $\mathord{\lesssim} 20\, \mathrm{ms}$ \citep{lyne_15}. Such cases can
still be explained by invoking sufficiently rapid rotation of
the progenitor core or spin-up by fallback. Based on the correlation
between $J_\mathrm{PNS}$ and $v_\mathrm{kick}$ in our models, one
would not expect to find long birth periods for neutron stars
with high kicks. A number of the pulsars in recent observational
studies of kicks and spins \citep{noutsos_13}, however, do not conform
to this trend. At this stage, it is unclear whether this issue has 
its origin in our limited simulations of progenitor models or if it
can (partly) be explained by observational selection effects.
Evidently, the scatter of the emerging
correlation needs to be investigated with more simulations before we
can draw conclusions. Moreover, it is still unclear how spin-up or 
spin-down during the explosion affects rotating progenitors, and
whether the correlation seen in our models still holds in this case.
Another simple explanation could be that a fraction of neutron stars 
come from higher-mass progenitors that have much more complicated
explosion  dynamics. It is also noteworthy that the observed pulsars
with high kicks and long birth spin periods tend to be 
old ($\mathord{\gtrsim} 10^6\,\mathrm{yr}$), and on these time scales
magnetic field dissipation may become relevant \citep{pons_07a,pons_07b},
which could induce uncertainties in the inferred natal spin periods.

\subsubsection{Absence of Spin-Kick Alignment}
Observations have suggested a tendency towards spin-kick alignment
in young pulsars (\citealp{johnston_05,ng_07,noutsos_12,noutsos_13};
but see also \citealt{kaplan_08} for difficulties in determining the
angle between the spin axis and the kick velocity). 

Although various hypotheses have been formulated to explain this putative
finding  (see \citealp{spruit_98,lai_01,wang_07} and especially 
\citealp{janka_17} for a more exhaustive summary of ideas), 
hydrodynamical simulations have not borne out these ideas so far. No 
indication of spin-kick alignment was seen in the parameterised 3D models
of \citet{wongwathanarat_13} and \citet{gessner_18}. In their $18 M_\odot$
model, \citet{mueller_17} also obtained a large angle of $40^\circ$ 
between  the kick and spin direction at the end of the run, but found 
that this angle decreased steadily over time scales of seconds. They
speculated that this decrease might be due to a ``righting'' mechanism
from the preferential accretion of material at directions perpendicular
to the kick.

Our larger set of models does not support such a righting mechanism.
As shown by Figure~\ref{fig:alignment}, the angle $\alpha$ between
the spin and kick direction varies considerably in our models. If 
anything, the spin vectors cluster at $\alpha >50^\circ$, although
this should not be overinterpreted considering the small sample size.
There is no systematic trend towards a decrease of $\alpha$ due
to ongoing accretion; some models actually show an increase in
$\alpha$ at late times.

If there is a mechanism for spin-kick alignment, we have clearly
not identified it in 3D simulations of neutrino-driven explosion models
yet. Again, the impact of moderate progenitor rotation on the dynamics
of the explosion phase and the evolution of the PNS
spin and kick clearly needs to be investigated as an obvious
missing factor in the current models.

\section{Conclusions}
\label{sec:conclusions}

In this paper, we have presented a suite of 3D supernova  
models of low-mass hydrogen-rich and (ultra)-stripped-envelope progenitors
obtained with the \textsc{CoCoNuT-FMT} code \citep{mueller_15a}.
This allowed us to study the distribution of explosion and
remnant properties in this mass range and possible correlations
among them using self-consistent 3D long-time simulations for
the first time. We consider single star models of
$9.6 M_\odot$, $11.8 M_\odot$, $12 M_\odot$, and $12.5 M_\odot$,
and stripped binary star models with initial helium star 
masses of $2.8 M_\odot$, $3 M_\odot$, and $3.5 M_\odot$.
In two cases ($2.8 M_\odot$ and $3.5 M_\odot$) the subsequent 
binary evolution after the removal of the hydrogen envelope 
was followed further to stellar death as ultra-stripped
progenitors \citep{tauris_15}.

All of our models explode successfully by the neutrino-driven mechanism. Thanks to the relatively
fast drop of the mass accretion rate, shock revival occurs early
in most of these low-mass models, but we also find an example
of a late explosion about $0.5\, \mathrm{s}$ after bounce for
a $12.5M_\odot$ single-star progenitor. There is considerable 
variation in the explosion energies and the mass of iron-group
ejecta, which can be taken as a rough proxy for the nickel
mass. Explosion energies range from $10^{50}\, \mathrm{erg}$
to $4\times 10^{50}\, \mathrm{erg}$, although 
considerable growth may still occur in some of our models beyond
the simulated time. The mass of iron group ejecta falls between
$0.01 M_\odot$ and $0.04 M_\odot$. These values are compatible
with the more modest explosion energies and nickel masses
observed for hydrogen-rich low-mass progenitors \citep{pejcha_15c} 
as well as for the ultra-stripped supernova candidates
SN~2005ek \citep{drout_13,tauris_13}, SN~2010X \citep{kasliwal_10,moriya_17} and iPTF~14gqr \citep{de_18}.
Thus, neutrino-driven explosion models do not appear to be underenergetic compared to observations
in this mass range. The detailed comparison of some of the
simulations suggests that 3D explosion models retain some degree
of stochasticity in the explosion properties due to the complex
dynamics of outflows and downflows in the explosion phase.
For example, in the $11.8 M_\odot$ model the increase of the explosion
energy is slowed down by a reconfiguration of the outflow
geometry at about $1\, \mathrm{s}$. This bears some vague 
resemblance to the phenomenon of outflow quenching in 2D models 
\citep{mueller_15b}, although these stochastic flow variations
still have a much less dramatic impact than in 3D than in 2D.

This element of stochasticity and the small sample size
preclude any conclusions on systematic differences in
explosion and remnant properties between
single- and binary-star progenitors. At present, we have
no evidence that single- and binary-star models of
neutrino-driven explosion with similar helium core masses differ
more than expected from the stochastic variations among single-star
models alone. Nonetheless, binary mass transfer of course
remains a crucial factor in the evolution
of supernova progenitors, since it affects
the \textit{distribution} of key structural parameters
like the helium core mass
\citep[e.g.,][]{podsiadlowski_04} and -- via
the envelope structure -- mixing processes during the later
phases of the explosion and the observable transients.

For most of our models, we can already determine, or at least 
extrapolate, the final neutron star properties quite well, barring
the possibility of late-time fallback. Except for the $12.5 M_\odot$
model, mass outflow already dominates over mass accretion onto
the PNS, and the PNS mass has
practically stabilised at its final value. Correcting for
the binding energy of the neutron stars, we obtain gravitational
masses between $1.22 M_\odot$ and $1.44 M_\odot$, which is compatible
with the distribution of observed neutron star masses
\citep{oezel_16,antoniadis_16,tauris_17}. While the neutron star kicks
are still growing at the end of the simulations due to the long-range
gravitational tug by the asymmetric ejecta, the subsequent
acceleration of the neutron star can be smoothly extrapolated
to obtain tentative final values in all but one case.  The
extrapolated kicks range from $11 \, \mathrm{km}\, \mathrm{s}^{-1}$
to $695 \, \mathrm{km}\, \mathrm{s}^{-1}$.
Thus, the most extreme, ECSN-like models
with the smallest helium cores
can reproduce the very low kicks required
to explain some double neutron star
systems and pulsars in globular
clusters \citep{tauris_17}, while
the models with higher He
core masses are compatible
with the typical kicks of young pulsars 
\citep{arzoumanian_02,hobbs_05,ng_07}.
 If the extrapolated kick
of $1236 \,\mathrm{km}\,\mathrm{s}^{-1}$ for the
$18 M_\odot$ model of \citet{mueller_17} is included, the
3D \textsc{CoCoNuT-FMT} models roughly span the full range
of observed kick velocities. We see tentative evidence for
a correlation of the kick velocity with the explosion energy
as proposed by \citet{janka_17} and \citet{vigna_gomez_18} (as a refinement
of earlier ideas for progenitor-dependent kicks by 
\citealt{bray_16}). Our models suggest that this correlation
may not be a tight one, however, and that the kicks may
scatter between zero and an upper limit that scales with
the explosion energy. Low kicks can be achieved in more energetic
explosions if the explosion geometry is bipolar rather than
unipolar, as has already been noted in 2D by \citet{scheck_06}.
Such a bipolar explosion occurs in one of our seven
simulations  (the $12 M_\odot$ model). Although there
is some concern that the bipolarity may be connected to the
grid geometry, we find unipolar models even in cases where we
we do not include strong aspherical seed perturbations in
the convective O shell that break grid alignment; this suggests
that the possibility of bipolar neutrino-driven explosions with low 
kicks is real in 3D. We also find a loose correlation between
the neutron star mass and the kick velocity, which is
in line with current observations, and partly theoretical expectations, of double neutron stars \citep{tauris_17},
but cannot make as strong a case for this correlation based on our
simulations. An investigation of a larger suite of supernova simulations of ultra-stripped stars 
is needed to confirm this hypothesis.

Computing the spin-up of our non-rotating progenitor models
by asymmetric accretion during the supernova, we find a range
of neutron star birth spin periods from $2.749 \, \mathrm{s}$ down to
$20 \, \mathrm{ms}$. Again, the range of spin periods is compatible
with observational constraints of young radio pulsars 
\citep{faucher_06,perna_08,popov_12,noutsos_13,igoshev_13} even without
assuming any rotation in the progenitor. This underscores that
neutron star birth spin periods are at least as much determined by the spin-up
during the supernova itself as by the spin of the progenitor
cores. Although this does not render the question of the rotational
state of the progenitor cores irrelevant by any means, it poses
an obstacle for using neutron star spins as a probe of the
intricate problem of angular momentum transport by (magneto-)hydrodynamical 
processes in the interiors of massive stars \citep{heger_00,maeder_00,heger_05,fuller_15}.

As in parameterised 3D simulations \citep{wongwathanarat_13,gessner_18}
we do not find any evidence for spin-kick alignment.
If spin-kick alignment is indeed prevalent in young neutron stars
as suggested by observations \citep{johnston_05,ng_07,noutsos_12,noutsos_13}, some ingredient
is still missing in the current 3D models. It is possible that
the situation will change when rotation in the progenitors
is included, or that other mechanisms such as spin-kick
alignment in SASI-driven explosions or non-hydrodynamical
mechanisms are needed (see \citealp{janka_17} for current scenarios).
On the other hand, it is noteworthy that
\citet{bray_16} found no preference for spin-kick alignment in their
binary population synthesis study. Neither did \citet{tauris_17} find any such correlation 
based on simulations of the kinematic effects of the second supernova in known double neutron star systems. 
Close interaction between observations, kinematic studies and computational modelling
is called for to better address the question of spin-kick alignment.

Instead of spin-kick alignment, we find a correlation between
the spin frequency and the magnitude of the kick. This is in line
with theoretical expectations \citep{spruit_98}, but needs to
be squared with the well-established findings of rapidly spinning 
pulsars with  low kicks -- such as the Crab pulsar 
\citep{kaplan_08,hester_08}. In these cases, a sufficiently rapid
rotation rate of the progenitor core or late-time fallback would
still provide a simple explanation. Long birth spin periods
in combination with high kicks would provide a more serious challenge
to the current neutrino-driven models. Such a constellation has
been inferred for some pulsars \citep{noutsos_13}, although these
are sufficiently old that the applied method to infer their birth spin periods 
comes with some uncertainty.
Again, close interaction between observations and theory is required
to determine whether the neutrino-driven models are compatible
with the observational evidence.

It is clear that our simulations are only a first step towards
understanding the distribution of supernova explosion and remnant properties
by means of self-consistent 3D simulations. Work is still
required on many fronts. While we can already obtain reasonably safe
values for the final neutron star properties (albeit at the cost
of a physically motivated extrapolation in case of the kicks),
longer simulations are needed to obtain converged values
for  explosion energies and nickel masses. Especially as far
as the nucleosynthesis is concerned, the approximate nature
of our neutrino transport is also an issue; long-time
models with more sophisticated transport and full, state-of-the-art
neutrino interaction rates will be needed in the future. Rotating
progenitors have yet to be explored by means of self-consistent
long-time simulations, and a broader coverage of progenitor
masses is called for.  While our results for low-mass progenitors
with modest explosion energies are encouraging, we still need
to address progenitors with somewhat higher masses and more typical 
explosion energies. Our work demonstrates, however, that 
self-consistent 3D simulations are now in a position to 
explore the distribution of supernova explosion and remnant 
properties in a systematic way and link up with observations of 
transients and compact remnants.

\section*{Acknowledgements}
This work was supported by the Australian Research Council through
ARC Future Fellowships FT160100035 (BM), Future Fellowship
FT120100363 (AH), and through the
Centre of Excellence for Gravitational
Wave Discovery (OzGrav) under project number CE170100004 (JP), 
by STFC grant ST/P000312/1 (BM),
and by the US Department of Energy through grant DE-FG02-87ER40328
(YZQ). CC  was supported by an Australian Government Research Training Program (RTP) Scholarship. PB was supported in part by the National Natural Science Foundation of China Fund No. 11533006. This material is based upon work supported by the National Science Foundation under Grant No. PHY-1430152 (JINA Center for the Evolution of the Elements). 

This research was undertaken with the assistance of
resources obtained via NCMAS and ASTAC  from the National Computational Infrastructure (NCI), which
is supported by the Australian Government and was supported by
resources provided by the Pawsey Supercomputing Centre with funding
from the Australian Government and the Government of Western
Australia.  This work used the DiRAC Data Centric system at Durham
University, operated by the Institute for Computational Cosmology on
behalf of the STFC DiRAC HPC Facility (\url{www.dirac.ac.uk}); this
equipment was funded by a BIS National E-infrastructure capital grant
ST/K00042X/1, STFC capital grant ST/K00087X/1, DiRAC Operations grant
ST/K003267/1 and Durham University. DiRAC is part of the UK National
E-Infrastructure. The authors acknowledge the Minnesota Supercomputing
Institute (MSI) at the University of Minnesota for providing resources
that contributed to the research results reported within this paper.

\appendix

\section{Mesh Coarsening and Filtering}
\label{sec:filter}
Spherical polar coordinates are highly useful for core-collapse
supernova simulations due to the approximate spherical symmetry of the
flow in many regions and the pronounced radial stratification over
many orders of magnitude in density.  They suffer from one major disadvantage,
however, in that the convergence of the grid lines near the polar axis
severely constrains the time step.  Aside from giving up the spherical
coordinate geometry altogether in favour of Cartesian geometry and
adaptive mesh refinement, various approaches have been applied to
alleviate the time step constraints near the axis without sacrificing
the advantages of spherical geometry. Overset orthogonal grids
\citep{kageyama_04,wongwathanarat_10a} and non-orthogonal grids
\citep{wongwathanarat_16} are excellent solutions, but add some
complexity to the grid geometry. An alternative approach is to retain
the basic structure of a spherical polar grid, but to adaptively
combine (``coarsen'') cells to at high latitudes and (optionally) also
close to the origin. Such an approach has been implemented by
\citet{mueller_15b}, and in a somewhat different way by
\citet{skinner_18} as a ``dendritic grid''.  Yet another alternative
consists in filtering the solution in Fourier space, which has a long
tradition in meteorology \citep{boyd}.

The \textsc{CoCoNuT} code offers both mesh coarsening and filtering in
Fourier space as an option. Since the implementation of these methods in
\textsc{CoCoNuT} has not been described in detail before, we here
provide a brief sketch.

\subsection{Mesh Coarsening}
In the case of mesh coarsening, we combine several zones in the
$\varphi$-direction at high latitude to larger ``supercells''. The
resolution $\Delta \varphi_\mathrm{SC}$ of the supercells is given by
\begin{equation}
    \Delta \varphi_\mathrm{SC}=
    2^{[-\log_2 \sin\theta]} \Delta \varphi ,
\end{equation}
where square brackets denote the floor function. This prescription ensures that
the time step constraint is only about a factor two worse at high latitudes than at the equator.

All the conserved and primitives variables remain defined on the
entire fine grid, and the conserved variables are updated in the usual
manner using the Riemann fluxes.  Before the recovery of the primitive
variables, we first average all the conserved variables in each
supercell, i.e.\ in the relativistic case, the Eulerian baryonic mass
density $D$, the components $S_r$, $S_\theta$, and $S_\varphi$ of the
relativistic momentum density in the spherical polar basis, the
Eulerian energy density $\tau$, as well as $D X_i$ for the mass
fractions $X$ of all species. To recover at least second-order
accuracy, we then prolongate the averaged values in the supercells
back to the fine grid using piecewise linear reconstruction. In
principle, higher-order conservative reconstruction can be used for
this purpose, but we opt for piecewise linear reconstruction
(typically using van Leer's harmonic limiter;
\citealp{van_leer_74}). One reason for this is that the parallelization
of slope-limited linear reconstruction in supercells is simpler.

There is also a more subtle reason for piecewise linear
reconstruction, however. It turns out that it is necessary to forgo
reconstruction of the conserved variables, because this can easily
lead to unphysical thermodynamic conditions in fine-grid cells after
prolongation if the shapes of the interpolants of the different
variables do not match in a reasonable manner. To avoid this problem,
we instead use linear reconstruction for $D$, $S_i/D$, $\tau/D$, and
$X_i$, which effectively ensures that the primitive variables obtain
reasonable values. However, this necessitates the following correction
procedure to make the interpolation conservative: If we let $S_Y$
denote the limited slope for variable $Y$, we reconstruct the
conserved variable $D Y$ as
\begin{equation}
    (D Y) (\varphi_j')
    =\overline{D Y} +
    D (\varphi_j') S_Y
    \left(\varphi_j'-\varphi_{\textsc{SC},j}-
    \frac{S_D \,\Delta \varphi_\mathrm{SC}}{12 \bar{D}}\right),
\end{equation}
where $\varphi_j'$ and $\varphi_{\textsc{SC},j}$ are the
$\varphi$-coordinates of the fine cell and supercell centres, and
barred quantities are supercell averages. Standard piecewise linear
conservative reconstruction is used for $D$. This correction procedure
turns the interpolants for the conserved variables $S_i$, $\tau/D$,
and $DX_i$ into quadratic functions and ensures that the scheme remains
conservative. Furthermore, we apply additional limiting to the slopes
of the ``quotient'' variables,
\begin{equation}
    S_Y\rightarrow \frac{S_Y}{1+|S_D \,\Delta \varphi_\mathrm{SC}/12 \bar{D}|},
\end{equation}
which avoids spurious overshooting of the interpolants due
to the above correction procedure if the slope in $D$ is very steep.

The prolongated solution on the fine grid is then used for the
recovery of the primitive variables, for PPM reconstruction and the
computation of the fluxes, and any other operations other such as the
solution of the transport equation.

\subsection{Filtering in Fourier Space}
In the newer, alternative scheme we filter the conserved variables in
Fourier space after every update. After computing the FFT in the
$\varphi$-direction, we suppress high-wavenumber modes in the Fourier
transform $\tilde{Y}(k)$ of any conserved quantity by multiplying with
a filter function
\begin{equation}
\tilde{Y}(k)
\rightarrow
\tilde{Y}(k)
\min
\left[\frac{r \,\Delta \varphi \sin \theta}{2 R_0\,\Delta \theta 
\sin (k \, \Delta\varphi/2)},
1
\right],
\end{equation}
where $k$ is the wavenumber in Fourier space, and $R_0$ is the radius
of the spherical inner core. The form of the filter function is
inspired by typical choices in meteorology \citep{jablonowski_11}, with
the important distinction that we filter the conserved variables and
not the fluxes as is often done in numerical weather prediction and
climate modelling. This filtering procedure very quickly damps
unstable modes with wavelengths shorter than $r \Delta \theta/2$. This
implies that the allowed Courant time step at high latitudes is
roughly half as long as at the equator.

One obvious advantage of filtering in Fourier space is that the filter
is gradually pushed to higher wavenumbers and eventually switched off
for larger radii without the need to implement a complicated data
layout and MPI communication pattern. A potential problem with
filtering in Fourier space is the occurrence of Gibbs phenomena, but
this does not seem to be of relevance in practice for two reasons. The
gradual decrease of the damping factor helps to avoid spurious
oscillations, and, furthermore, the radial dependence of the damping
term ensure that little filtering is done at the shock radius, where
the Gibbs phenomenon would present the most serious problem.

\bibliography{paper}

\label{lastpage}

\end{document}